\documentclass[nofootinbib,nobibnotes,bibnotes,amsmath,amssymb,floatfix,aps,pra,twocolumn,10pt]{revtex4-2} 

\usepackage{graphicx}
\usepackage{subcaption}
\usepackage{adjustbox}
\usepackage{dcolumn}
\usepackage{bm}
\usepackage{hyperref}
\usepackage[mathlines]{lineno}
\usepackage{soul,xcolor} 

\usepackage{placeins}
\usepackage{pifont}
\newcommand{\xmark}{\ding{54}}%

\usepackage{enumitem}
\setlist[enumerate]{leftmargin=*, 
                    }

\begin{document}

\preprint{APS/123-QED}

\title{Emergent Microrobotic Behavior of Active Flexicles in Complex Environments}

\author{Sophie Y. Lee}
    \thanks{These two authors contributed equally}
\affiliation{%
Department of Materials Science and Engineering, University of Michigan, Ann Arbor, Michigan 48109, USA
}%

\author{Philipp W. A. Sch\"{o}nh\"{o}fer}
    \thanks{These two authors contributed equally}
\affiliation{
  Department of Chemical Engineering, University of Michigan, Ann Arbor, Michigan 48109, USA
}%

\author{Sharon C. Glotzer}
  \altaffiliation{corresponding author}
\affiliation{%
Department of Materials Science and Engineering, University of Michigan, Ann Arbor, Michigan 48109, USA
}%
\affiliation{%
Department of Chemical Engineering, University of Michigan, Ann Arbor, Michigan 48109, USA
}%
\affiliation{%
Biointerfaces Institute, University of Michigan, Ann Arbor, Michigan 48109, USA
}%

\date{\today}

\begin{abstract}
\noindent Collections of simple, self-propelled colloidal particles exhibit complex, emergent dynamical behavior, with promising applications in microrobotics. When confined within a deformable vesicle, self-propelled rods cluster and align, propelling the vesicle and inducing changes in the vesicle shape. We explore potential microrobotic capabilities of such vesicle-encapsulated particles, which form a composite particle system termed a `flexicle'. Using molecular dynamics simulations, we demonstrate that the alignment of rods enables flexicles to locomote and respond adaptively to their physical environment. When encountering solid boundaries or obstacles, the rods reorient at the interface, triggering novel emergent behaviors such as crawling, corner-preferencing, wall climbing, and object-latching. These interactions and accompanying internal rod re-arrangement lead to spontaneous, temporary differentiation of the rods into `latchers' and `navigators'. This division of labor among the rods enables coordinated locomotion and environmental response. Our findings establish flexicles as a versatile platform for programmable, geometry-sensitive microrobotic behavior, offering a step toward autonomous colloidal robotics.
  
\end{abstract}

\maketitle

\section{\label{sec:intro}Introduction}

Robots are becoming increasingly ubiquitous, from robotic arms on assembly lines to autonomous vehicles in warehouses and on the roads, to self-driving vacuum cleaners and robots assisting in surgeries. Thus, it is no surprise that researchers have dedicated substantial effort to designing and creating intelligent colloidal particles to initiate a comparable revolution on the microscale~\cite{liu_colloidal_2023,bishop_active_2023}. In parallel with the intensified interest in synthetic active matter, in which constituent particles convert external energy into motion, researchers have proposed a diverse set of methods for efficient and precise control of these ``self-propelled" particles (SPPs) and their swarms. Such strategies range from purely computational and mathematical models~\cite{vicsek_novel_1995,ballerini_interaction_2008,schaller_polar_2010,palacci_living_2013,deseigne_vibrated_2012} to experimental realizations involving both biological~\cite{gumuskaya_motile_2023,park_directed_2016,lushi_collective_2012,reversat_cellular_2020,keya_dna-assisted_2018} and synthetic microscopic systems~\cite{li_development_2018,solovev_magnetic_2010,nelson_micro-_2014,gardi_microrobot_2022,palacci_living_2013,akter_cooperative_2022}.\\

To date, most active microparticles made in the lab depend on external mechanisms for guidance --- such as magnetic fields~\cite{zhang_artificial_2009,ghosh_controlled_2009,tierno_controlled_2008,akolpoglu_magnetically_2022,huang_soft_2016}, light~\cite{jiang_active_2010,debata_light-driven_2024,palagi_lightcontrolled_2019,palacci_living_2013} or acoustic sources~\cite{collins_acoustic_2016,ahmed_artificial_2016,wang_autonomous_2012,qiu_swimming_2014,garciagradilla_ultrasoundpropelled_2014}. These external fields enable active agents to adapt to changing environments or to perform specific robotic tasks such as navigation~\cite{dong_highly_2016,li_magnetically_2016} and transportation of objects~\cite{demirors_active_2018,ghosh_controlled_2009,dong_highly_2016,garciagradilla_ultrasoundpropelled_2014}. Although the design space available for active particle synthesis includes particle shape~\cite{lum_shape-programmable_2016,ren_multi-functional_2019,ebbens_pursuit_2010,venugopalan_fantastic_2020,gao_cargotowing_2012}, propulsion mechanism~\cite{sundararajan_catalytic_2008,paxton_chemical_2006,sanchez_cilia-like_2011,gao_synthetic_2014}, and electromagnetic properties~\cite{hawkes_programmable_2010,zhou_magnetically_2021,servant_controlled_2015,zhang_artificial_2009}, current artificial microparticle systems lack autonomous mechanisms that can adjust the dynamics of a single particle in response to external triggers. Only recently have colloidal particle ``robots" been equipped with electronic devices and computing units that enable simple communication and response mechanisms to be programmed between particles~\cite{chen_bioinspired_2018,miskin_electronically_2020,palacci_living_2013}. In combination with advances in reinforcement learning models~\cite{falk_learning_2021,cichos_machine_2020}, these particles have the potential to eventually lead to even more complicated single-particle navigation strategies. However, significant challenges remain in colloidal electronics, particularly in improving the robustness and minimizing power consumption~\cite{miskin_electronically_2020}.\\

An alternative approach to pre-programming colloidal particles for autonomous behaviors involves bioinspired designs that mimic hierarchical concepts found in the microcosm and rely on the emergence of robotic functions. Single-celled organisms, for instance, migrate through coordinated interactions of intracellular active components, such as the cytoskeletal actin network, which exerts collective forces on the cell membrane~\cite{pollard_actin_2009}. Beyond propulsion, the cytoskeletal network also plays a significant role in sensing and adaptive response mechanisms, where environmental stimuli induce internal rearrangements that alter the cell's dynamical behavior~\cite{fletcher_cell_2010,cheng_roles_2020}. In this context, macroscopic robotic systems with hierarchical designs have been suggested~\cite{boudet_collections_2021,savoie_robot_2019,veenstra_adaptive_2025}. Furthermore, previous experimental studies have explored active synthetic Janus particles~\cite{vutukuri_active_2020} and motile bacteria~\cite{le_nagard_encapsulated_2022,ramos_bacteria_2020} encapsulated within vesicles. However, the latter systems primarily exhibit deformations of the vesicle rather than directed motion of the vesicle, as the encapsulated particles do not form polar clusters capable of directional swarm motion, and/or the low bending modulus of the membrane leads to tether formation~\cite{heinrich1999vesicle}. 

Recently, a model for a synthetic self-driven microparticle system that uses an analogous hierarchy mechanism for adaptive behavior is the ``flexicle''~\cite{schonhofer_collective_2025}. Flexicles are 3-dimensional, deformable, complex particle systems comprised of a collection of nanometer-to-micron sized colloidal particles suspended in a fluid and encapsulated by a flexible membrane, or vesicle. When the encapsulated particles are SPPs, their clustering ability couples to the membrane elasticity to deform and propel the flexicle. It was recently shown~\cite{schonhofer_collective_2025} that these superstructures, along with their 2D counterparts~\cite{lee_complex_2023,boudet_collections_2021,paoluzzi_shape_2016}, are actuated by an emergent cluster of aligned internal self-propelled rods that push perpendicularly against the vesicle boundary. This internal organization enables flexicles to function as single active agents, demonstrating adaptive responses to compression through dynamic reconfiguration of their internal components.\\ 

Here, we present simulations that demonstrate how the interplay between internal rod configurations and membrane deformability enables an array of robotic functionalities through emergence when flexicles encounter rigid and non-rigid obstacles. We demonstrate that flexicles can attach to spherical and cylindrical objects as their internal rods align against these objects. The dynamic reorganization of the internal rods and their spontaneous division into ``navigators'' and ``latchers'' --- actuators performing different tasks --- produces a wide range of robotic behaviors, including orbiting around obstacles, transporting cargo, and overcoming topographic barriers.  
\section{\label{sec:method}Method}

\begin{figure}[t!]
    \includegraphics[width = 0.45 \textwidth]{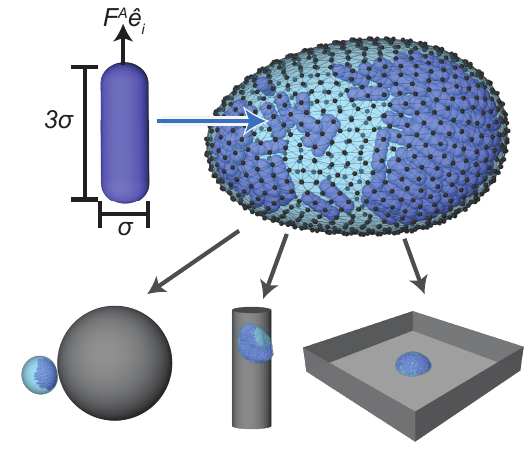}
    \caption{\label{fig:method} Illustration of the flexicle model considered in this paper. The top left snapshot depicts a self-propelled rod particle. The spherically capped rod is modeled as five closely overlapping spheres with diameter $\sigma$ rigidly connected along their diameters. The top right image shows a flexicle with 153 rod particles enclosed within a deformable membrane, which is represented by vertices connected by a triangulated mesh. The three images on the bottom represent the different external geometries that flexicles encounter in our study: a sphere, a cylinder, and a square box with walls.}
\end{figure}
  
We conducted molecular dynamics simulations of $N_r = 153 $ self-propelled rod-shaped particles with thickness $\sigma$ and length $3\,\sigma$ confined within a deformable vesicle of radius $R_\text{flex}=8~\sigma$ inside a cubic simulation box with side length $L$ and periodic boundary conditions. The vesicle membrane is represented by a triangulated mesh, which consists of $N_v$ = 900 vertices bonded to nearest neighbor vertices through the edges of the mesh. The rod-shaped particles are modeled as rigid multi-sphere bodies composed of five overlapping spherical beads connected linearly (see Fig.~\ref{fig:method}). 
Both the mesh vertices that comprise the membrane,

\begin{align}
m\,\dot{\mathbf{v}_i} &= \sum_{j}\mathbf{F}_{ij}^\text{WCA} + \mathbf{F}_i^\text{mesh}+ \mathbf{F}^\text{ext} -\gamma_{m}\,\mathbf{v}_i + \mathbf{F}^R
\end{align}

and the self-propelled rods,

\begin{align}
\label{eq:rod_langevin}
m\,\dot{\mathbf{v}_i} &= \sum_{j}\mathbf{F}_{ij}^\text{WCA} + F_{i}^{A}\hat{\mathbf{e}}_i -\gamma_a\,\mathbf{v}_i + \mathbf{F}_{i}^R\\
I\,\dot{\bm{\omega}_i} &= \sum_{j}\mathbf{T}_{ij}^\text{WCA} - \gamma_{i,r}\,\bm{\omega}_i + \mathbf{T}_{i}^R
\end{align}
  
follow Langevin dynamics~\cite{schlick_molecular_2010} with mass $m$ and moment of inertia $I$. We assigned translational drag coefficients $\gamma_m = 5~m\,\tau^{-1}$ for mesh nodes and $\gamma_a = 250~m\,\tau^{-1}$ for self-propelling rods, where the unit time is defined as $\tau = \sqrt{\frac{m\,\sigma^2}{\epsilon}}$ with $\epsilon=1~k_B\,T$. The rotational drag coefficient for the rods, $\gamma_{r}$, was calculated as $\frac{\sigma^2\,\gamma_{a}}{3}$ according to the Stokes-Einstein relationship~\cite{von_smoluchowski_zur_1906}.\\
  
Brownian forces $\mathbf{F}_{i}^R$ and torques $\mathbf{T}_{i}^R$ on the rods are modeled as $\mathbf{F}_{i}^R = \sqrt{2\,m\,\gamma_i\, k_B\,T}\,\eta_i(t)$ and $\mathbf{T}_{i}^R = \sqrt{2\,I\,\gamma_{r,i}\,k_B\,T}\,\zeta_i(t)$ with $\eta_i(t)$ and $\zeta_i(t)$ being normalized Gaussian white noise processes with zero mean ($\langle \eta_i(t) \rangle = 0$ and $\langle \zeta_i(t) \rangle = 0$)and unit variance ($\langle \eta_i(t)\,\eta_j(t') \rangle = \delta_{ij}\,\delta(t-t')$ and $\langle \zeta_i(t)\,\zeta_j(t') \rangle = \delta_{ij}\,\delta(t-t')$). 
Interparticle forces --- between beads comprising the rods (bead-bead) and between each rod bead and mesh vertices (bead-mesh) --- were modeled with the purely repulsive Weeks-Chandler-Andersen (WCA) potential~\cite{weeks_role_1971}, from which the forces $\mathbf{F}{ij}^{\mathrm{WCA}}$ and torques $\mathbf{T}{ij}^{\mathrm{WCA}}$ on each rod bead were computed.
  
\begin{align}
\label{eq:WCA}
  U^\text{WCA}(r_{ij}) = 
  \begin{cases} 
    4~\epsilon\,[ 
    ( \frac{\sigma}{r_{ij}})^{12}
    - ( \frac{\sigma}{r_{ij}})^{6}
    ]+\epsilon & r_{ij} < r_c = \sqrt[6]{2}~\sigma \\
    0 & r_{ij} > r_c,
  \end{cases}
\end{align}
with particle distance $r_{ij}=||\mathbf{r}_i-\mathbf{r}_j||$. The active force term, $F_{i}^{A}\,\hat{\mathbf{e}}_i$, is applied only to the center of the rods and is oriented along the long axis of the rod, $\hat{\mathbf{e}}_i$. The magnitude of the active force, $F_i^{A}$ is given by the P\'eclet number, Pe = $\frac{F_i^{A}\cdot\sigma}{k_B\,T}$. To ensure that the internal rods can propel the flexicle we set $\text{Pe}=100$ or $\text{Pe}=200$.\\

The term $\mathbf{F}_i^\text{mesh}$ represents the mesh-specific forces applied to the edge-linked vertices given by the mesh potential:
\begin{align}
  U^\text{mesh} &= \sum_{j} U_B(r_{ij}) +\sum U_H +\sum U_A
\end{align}

Neighboring edge-linked mesh vertices are bonded by a potential~\cite{noguchi_dynamics_2005} that combines both attraction and repulsion: 
\begin{gather}
    U_B(r) = U_{\mathrm{att}}(r) + U_{\mathrm{rep}}(r) \\
  \begin{align*}
  U_{\mathrm{att}}(r)  =
  \begin{cases}
  \kappa_B \frac{\exp(1/(l_{c0}-r))}{l_{max}-r}  & r > l_{c0} \\
  0 & r \leq l_{c0} \\
  \end{cases}\\
  U_{\mathrm{rep}}(r)  =
  \begin{cases}
  \kappa_B \frac{\exp(1/(r-l_{c1}))}{r-l_{min}}  & r < l_{c1}\\
  0 & r \ge l_{c1} \\
  \end{cases}
  \end{align*}\\
\end{gather}
Here we define the bond stiffness $\kappa_B=1334~k_B\,T$, the maximally allowed edge length $l_{max}=1.34~\sigma$, the lower cutoff edge length for the attractive force $l_{c0}=1.15~\sigma$, the upper cutoff edge length for the repulsive force $l_{c1}=0.85~\sigma$ and the minimally allowed edge length$l_{min}=0.67~\sigma$.\\
  
The bending force between adjacent triangles in the mesh is calculated using the Helfrich curvature energy~\cite{helfrich_elastic_1973,gompper_random_1996}.
\begin{align}
  U_H &= 2~\kappa_H \oint_{S}{\mathbf{H}^2 \cdot d\mathbf{S}} \approx \sum_{i=1}^{N_v} \frac{2~\kappa_H\,H_i^2\,S}{Nv}\\
  &=\frac{\kappa_H}{2}\sum_{i=1}^{N_v} \frac{1}{s_i}[\mathbf{n}_i \cdot (\sum_{j(i)} \frac{d_{ij}}{r_{ij}}\,\mathbf{r}_{ij})^2],
\end{align}
where $\kappa_H$ represents the bending rigidity of the mesh, $H$ is the local mean curvature, $S$ is the total mesh area and $H_i$ is the mean curvature at vertex $i$. The normal unit vector of the mesh at the vertex $i$ is $\mathbf{n}_i$. The area of the dual cell at the vertex $i$ is given by $s_i=(\sum_{j(i)}d_{ij}\,r_{ij})/4$, where $j(i)$ represents all neighboring vertices connected to the vertex $i$. The length of the bond in the dual network is defined as $d_{ij}=r_{ij}\,(\cot\theta_1 +\cot\theta_2)/2$, where $\theta_1$ and $\theta_2$ are the angles at the vertices opposite to the vector of the shared bond $\mathbf{r}_{ij}$.\\
  
Additionally, the local area of each triangle within the mesh is maintained using a harmonic potential described by:
\begin{align}
  U_A = \frac{\kappa_A}{2}\sum_{i=1}^{N_t}\frac{(A_i -A_0)^2}{A_0},
\end{align}
where $\kappa_A=10000\,k_B\,T$. Here, $A_i$ represents the instantaneous area of the i-th triangle, and $A_0$ is the desired area for each triangle, calculated as $A_0 = \frac{S}{N_t} = \frac{4\pi \cdot R^3}{3\cdot N_t}$. The total number of triangles in the mesh is given by $N_t=2\,(N_v-2)$. In this way, the surface area of the mesh remains conserved but the volume of the flexicle can vary.\\

To model fluidity of the vesicle, we implemented edge flips between neighboring triangles in the mesh, allowing the topology of the triangulation to change dynamically~\cite{nelson_statistical_2004, gompper_network_1997}. The flip process is executed every $10^{-1}~\tau$ time steps and involves a Monte Carlo trial flip applied to each edge in the mesh. For each attempted flip of an edge $e$, we calculate the change in energy $\Delta U_e^\text{mesh} =U_\text{after} - U_\text{before}$ and accept the new edge with probability $\Psi=\text{min}(1,\text{exp}[-\Delta U_e^\text{mesh}/k_B\,T])$.\\

We use four different shapes for the external objects encountered by flexicles: a sphere, a cylinder, a square box with surrounding walls and a flight or stairs. These objects exert a steric force $\mathbf{F}_i^\text{ext}$ on the flexicle membrane upon contact. The spherical and cylindrical obstacles have radii $3~\sigma<R_\text{sph/cyl}<15~\sigma$. The contact force between obstacle and mesh vertex is modeled by a radially shifted WCA potential.
\begin{align}
\label{eq:S_WCA}
U^\text{SWCA}(\tilde{r}_{i}) = 
4~\epsilon [ 
( \frac{\sigma}{\tilde{r}_{i}-\Delta})^{12}
- ( \frac{\sigma}{\tilde{r}_{i}-\Delta})^{6}
]+\epsilon,
\end{align}
with $\Delta=R_\text{sph/cyl}$ and the distance $\tilde{r}_{i}=||\mathbf{r}_i-\mathbf{c}^\text{obst}||$ between the mesh vertex and the sphere center/cylinder symmetry axis $\mathbf{c}^\text{obst}$.\\

The square box $80~\sigma$ wide and $80~\sigma$ long consists of two elements: the floor and four surrounding walls. We model the floor as a smooth, flat interface in the xy plane, which interacts with the mesh vertices according to a Lennard-Jones (LJ) potential:
\begin{align}
U^\text{floor}(r_{ij}) = 
\begin{cases} 
  4~\epsilon [ 
  ( \frac{\sigma}{r_{ij}})^{12}
  - ( \frac{\sigma}{r_{ij}})^{6}
  ] & r_{ij} < r^\text{floor}_c\\
  0 & r_{ij} > r^\text{floor}_c
\end{cases}
\end{align}
To simulate adhesion between a flexicle and the floor, we set the cutoff distance $r^\text{floor}_c= 3\sqrt[6]{2}~\sigma$. For purely repulsive interfaces, we set $r^\text{floor}_c= \sqrt[6]{2}~\sigma$. The vertical side walls that bound the domain are constructed explicitly from $N_\text{wall}=7632, 12720$ and $ 17808$ static particles, corresponding to side-wall heights $h=6~\sigma$, $10~\sigma$, and $14~\sigma$, respectively. The interactions between the wall particles and the mesh vertices follow Eq.~\eqref{eq:WCA}. Lastly, the stairs consist of 5 steps, each being $8~\sigma$ high and $16~\sigma$ wide. Like the box walls, the steps are constructed from $N_\text{step}=10000$ static particles.\\

We ran each simulation at a temperature $k_B\,T=0.2$ for a time period $t=1\times10^5\tau$ with a time step $\Delta t=1\times10^{-3}~\tau$. We chose our parameter space for the bending rigidities $\kappa_H \in [5\,k_B\,T,10\,k_B\,T,100\,k_B\,T,500\,k_B\,T,1000\,k_B\,T,2000\,k_B\,T]$. In the Supplementary Information (SI), we provide a table~\ref{tab:units} that converts all parameters into experimental units using experimentally realistic base units. The time interval between each sampled snapshot is $100000~\Delta t$. We used the open-source molecular dynamics software \textsc{HOOMD-blue}~\cite{anderson_hoomd-blue_2020} [v4.0.0] to perform our simulations, the \textsc{Freud} data analysis package~\cite{ramasubramani_freud_2020} for cluster analysis, and the \textsc{signac} software package~\cite{adorf_simple_2018} for data management.
\section{\label{sec:results}Results}

When activating self-propelled rods enclosed within vesicles, we observe that most rods spontaneously agglomerate at the membrane interface and form a polar cluster, as observed in an earlier study of flexicles~\cite{schonhofer_collective_2025} and in a study of active rods inside polymer rings~\cite{abaurrea-velasco_vesicles_2019}. As the cluster forms, the rods induce pockets of high Gaussian curvature in the membrane and, owing to their anisotropic shape, orient themselves perpendicularly to the vesicle surface. Once a dominant cluster of aligned rods is formed, it generates sufficiently strong collective driving forces to push the entire flexicle forward, assuming that the (presumed) surrounding solvent can permeate through the membrane. This turns the flexicle into an active agent capable of exploring its environment. Although the pushing cluster remained stable inside every flexicle we simulated, the flexicle shape, as well as the direction and persistence of the flexicle motion, is governed by the membrane’s bending rigidity, the rods’ propulsion speed, and the number of rods inside the flexicle (see Fig.~\ref{fig:S1}). The velocity of the flexicle increases approximately proportionally with the propelling speed of the rods. However, as the rod density increases, the flexicle's velocity decreases (Fig.~\ref{fig:S1}(a)). This reduction is attributed to the formation of multiple rod clusters at higher densities, which generate competing forces that partially cancel each other, thereby reducing the net propulsion (see Fig.~\ref{fig:S1}(b)). Furthermore, the angle between successive displacement vectors of the flexicle reflects its rotational diffusion, which serves as an inverse measure of directional persistence. Rotational diffusion increases (and thus directional persistence decreases) with an increasing number of rods inside the flexicle up to $N_r=153$, after which it slightly decreases (Fig.~\ref{fig:S1}(c)). In the following, we focus on flexicles that reliably only form a single cluster within the studied bending rigidity window ($N_r=153$, $\text{Pe}=100,200$).

\subsection{\label{sec:sphere}Flexicle on a Sphere}

\begin{figure*}[t!]
\includegraphics[width = 0.9 \textwidth]{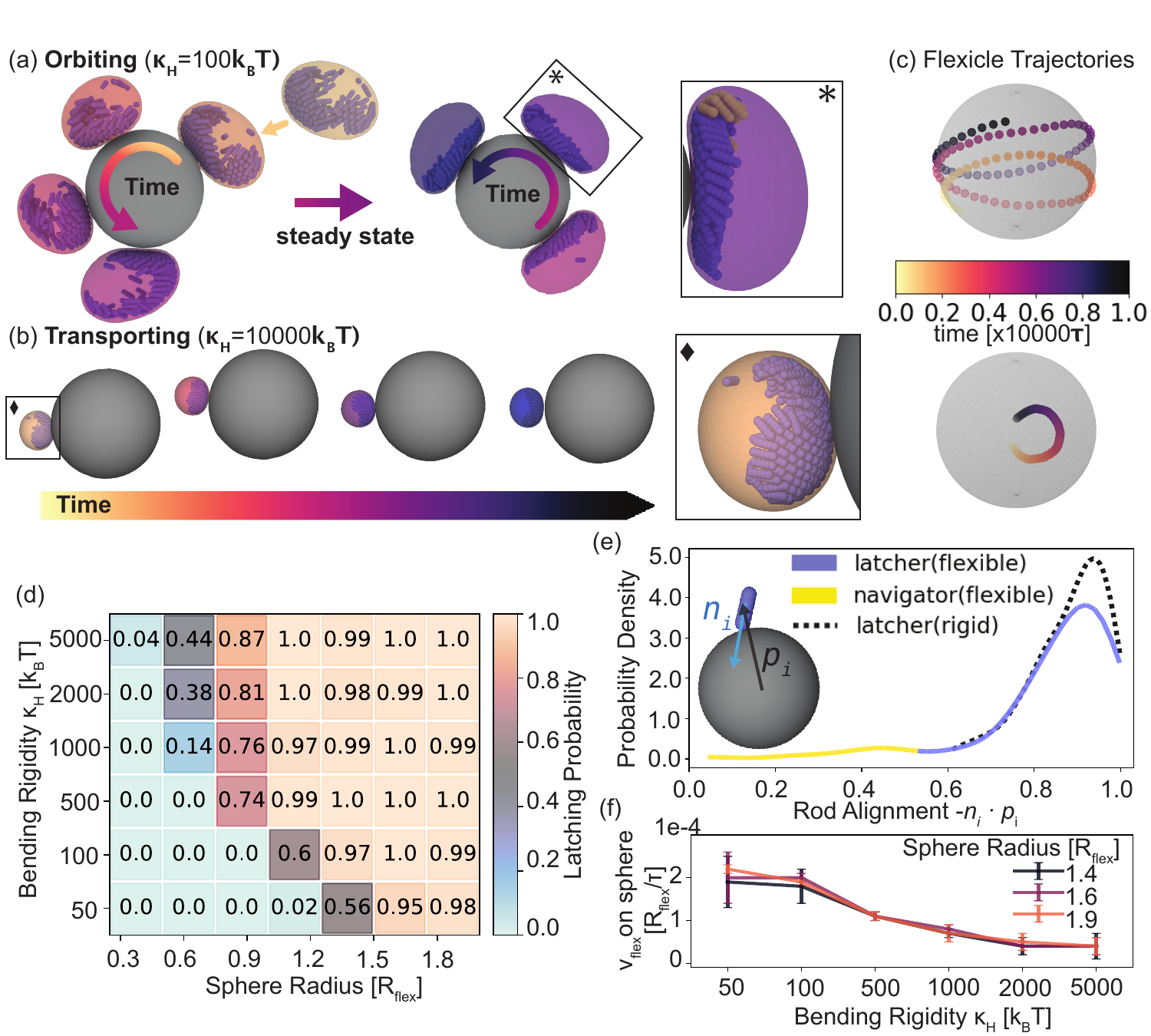}
\caption{\label{fig:results1} (a) Collision and latching behavior of a flexicle ($\kappa_H=100~k_B\,T$) interacting with a spherical obstacle ($R_\text{sph}=1.3~R_\text{flex}$). Left: Five snapshots show the latching process. Middle: Three snapshots illustrate steady-state circulation around the sphere. Right: Internal rod organization highlights two roles—latchers (blue) and navigators (yellow). (b) Four snapshots showing a rigid flexicle ($\kappa_H=10000~k_B\,T$) transporting a movable sphere in a straight line. Inset (far right): Rod alignment at the initial time point. (c) Trajectory over time of a flexicle moving on a static sphere. Top: Flexible membrane ($\kappa_H = 100~k_B\,T$). Bottom: Rigid membrane ($\kappa_H = 10000~k_B\,T$). (d) Latching probability as a function of obstacle radius.(e) Alignment distribution of rods, measured by the angle between each rod's axis ($n_i$) and the surface normal ($\hat{p}_i$). Solid and dotted lines show results for flexible and rigid membranes. Peaks are color-coded to match rod roles in (a). (f) Instantaneous velocity of flexicles on static spheres across varying membrane rigidities. All simulations were performed at Péclet number $\mathrm{Pe} = 100$.}
\end{figure*}

Upon encountering a static sphere, we observe that flexicles can latch onto the object and remain on the surface for an extended period of time; we refer to this as stable latching (see Fig.\ref{fig:results1}(a) and Movie.1). For spherical obstacles, the probability of stable latching is influenced by both the object's radius $R_\text{sph}$ and the membrane's bending rigidity $\kappa_H$ (see Fig.\ref{fig:results1}(c)). Although flexicles slow down upon collision with small spheres ($R_\text{sph}\approx0.2~R_\text{flex}$) and occasionally remain latched for some time, they do not latch permanently. Instead, the flexicles leave the small obstacle once they regain speed. We find that only above a minimum sphere radius $R_\text{min}$, which depends on the membrane bending rigidity, do the flexicles permanently latch onto and move along the sphere's surface. Further, the minimum sphere radius for stable latching is lower for rigid flexicles $R_\text{min}(\kappa_H=5000~k_B\,T)\approx0.6\,R_\text{flex}$ than for more flexible flexicles $R_\text{min}(\kappa_H=50~k_B\,T)\approx1.3\,R_\text{flex}$.\\

To understand the mechanism underlying the latching process, we analyze the dynamics of the internal colloidal rods. When the flexicle collides with the sphere, its internal propelling rod cluster reorganizes, slowing the flexicle movement. If the sphere obstacle is large enough $R_\text{sph}>R_\text{min}$, we observe that most of the rods respond to contact with the sphere's surface by spontaneously pointing towards its center, as illustrated in Fig.~\ref{fig:results1}(a) and the angle analysis in Fig.~\ref{fig:S2}(c) and Fig.~\ref{fig:S3}(c). This internal reorientation mirrors earlier findings for self-propelled, free particles at rigid interfaces, where their motility drives them to accumulate along the surface~\cite{boudet_collections_2021, paoluzzi_shape_2016}. As a result of the cluster rearrangement, the rods collectively exert pressure on the membrane, pressing it further against the sphere, until a steady state is reached. After a flexicle latches onto a spherical obstacle, it stays attached; detachment is observed only when the obstacle's radius is later reduced during the simulation. The detachment radius, $R_\text{detach} < R_\text{min}$, varies with the shrinking rate of the sphere (see Fig.~\ref{fig:S4}), and is consistently smaller than the minimum latching radius, indicating a hysteresis between attachment and detachment.\\ 

\begin{figure}[t!]
\includegraphics[width = 0.5 \textwidth]{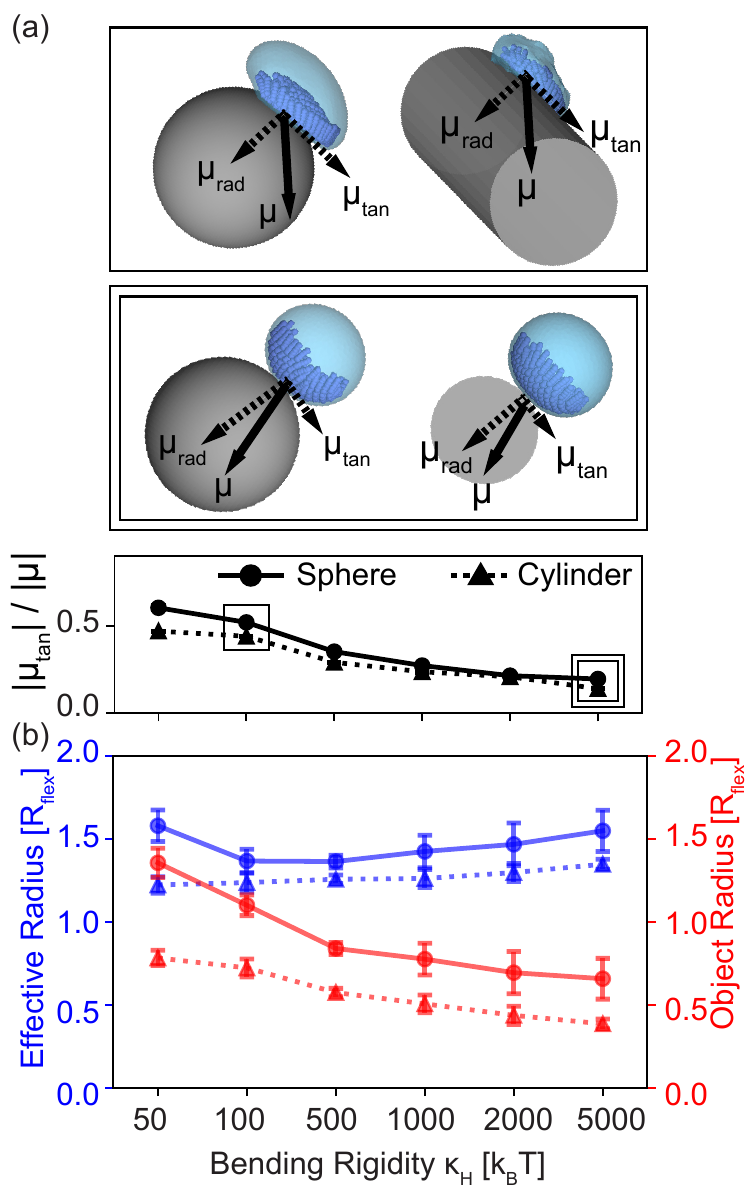}
\caption{\label{fig:results2} (a) Normalized tangential component of the net active force from the internal rod cluster as a function of membrane bending rigidity $\kappa_H$: $\frac{\mu_{\text{tan}}}{\sqrt{\mu_{\text{rad}}^2 + \mu_{\text{tan}}^2}}$, where $\mu_{\text{rad}}$ and $\mu_{\text{tan}}$ are the radial and tangential components of the net active force relative to the inward surface normal $\hat{\mathbf{n}}$. Solid lines represent results for spherical obstacles, and dotted lines for cylindrical ones. Insets show representative snapshots of flexicles interacting with the obstacles at $\kappa_H = 100~k_B\,T$ (top row) and $\kappa_H = 5000~k_B\,T$ (bottom row), with arrows indicating the direction of the net force. (b) Obstacle radius (red) and effective latching radius (blue) corresponding to a 50\% flexicle attachment probability, plotted against $\kappa_H$. Values are extracted from error-function fits to the latching probability data shown in Fig.\ref{fig:results1}(d) for spheres and Fig.\ref{fig:results3}(b) for cylinders. Solid and dotted lines follow the same notation as in (a). All simulations were performed at a Péclet number of $\mathrm{Pe} = 100$.}
\end{figure}

The latching process causes large shape deformations of flexicles with bending rigidities $\kappa_H \lessapprox 100~k_B\,T$. The flexicles become concave on the sphere-facing side, spreading and conforming to the surface curvature, and increasing the area of contact with the obstacle (see Fig.~\ref{fig:results1}(a) and Fig.~\ref{fig:S6}). Hence, the \textit{effective} radius of the obstacle, measured by the distance between the geometric centers of the sphere and flexicle, depends on the bending rigidity. Interestingly, we observe that the effective radius onset for stable latching is roughly universal across different bending rigidities (see Fig.~\ref{fig:results2}(b)).
The concave deformation of flexible flexicles produces pockets of high Gaussian curvature where the membrane bends away from the sphere. These high-curvature pockets align with the outer edge of the internal rod cluster, which occupies the entire contact zone between flexicle and sphere. Within this cluster, the internal rods exhibit a spatially heterogeneous organization. While rods in the cluster center are tightly packed and oriented predominantly toward the sphere centroid, rods at the boundary align along the curved membrane, bending away from the sphere surface. This difference in rod behavior is also evident in the angle distribution between the local surface normal and each rod orientation (see Fig.~\ref{fig:results1}(e)). The distribution reveals two distinct populations of rods, despite the fact that all rods are identical in type and propulsion strength. A dominant peak in the angle distribution corresponds to rods that align closely with the surface normal, while a secondary peak at larger angles reflects a subset of rods whose orientation is more tilted tangential to the surface. To highlight their distinct functional roles, we classify the rods from the first peak as \textit{latchers}, located primarily within the core of the cluster and contributing mainly to stable latching. In contrast, the rods of the secondary peak, which we label \textit{navigators}, are located at the front-facing pocket of high Gaussian curvature or the rear of the cluster. These navigators exert propulsion forces with significant tangential components (see Fig.~\ref{fig:results2}(a)), generating asymmetries in the collective force balance. As a result, they steer the flexicle along the spherical surface and drive orbital motion around the obstacle. This emergent, functional partitioning within the flexicle explains why more deformable vesicles exhibit a greater tendency to unlatch from the obstacle. Although these soft flexicles form larger contact area due to the membrane shape conforming to the sphere, the increased mobility and influence of navigators at the boundary enable dynamic reorientation and detachment.\\ 

As the bending rigidity increases, the contact area between the flexicle and the sphere decreases (see Fig.~\ref{fig:S6}) until the flexicles with $\kappa_H \gtrapprox 5000~k_B\,T$ touch the obstacle only at a single point, as shown in Fig.~\ref{fig:results1}(b) and Fig.~\ref{fig:S6}. In this regime, the flexicles remain spheroidal and indistinguishable from their unperturbed, free-moving state. With loss of contact area and concavity, the peak associated with navigators vanishes in the angle distribution of Fig.~\ref{fig:results1}(e). This indicates that none of the internal rods adopts tangential orientations. Instead, all rods form a coherent spherical cap aligned with the sphere normal, effectively functioning as latchers. The presence and absence of navigators signals a dynamic transition between deformable and rigid flexicle behavior, respectively. Deformable flexicles that retain navigators circumnavigate around the sphere obstacle along near–great-circle paths, whereas highly rigid flexicles, which lack navigators, either remain almost stationary or trace only small, localized loops on the sphere surface (see Fig.~\ref{fig:results1}(c)).\\ 

Based on the difference in dynamical behavior, we hypothesized that flexicles could be used for particle transport. To test this capability, we allow the sphere obstacle, previously held fixed, to move under Langevin dynamics similar to Eq.\ref{eq:rod_langevin} with a large mass $m_\text{sph}=700~m$. In this scenario, we observe that flexicles latch onto movable spheres in a manner consistent with the previously explained latching mechanism. For deformable flexicles, the presence of both latchers and navigators persists, causing the sphere to oscillate in space according to the orbiting frequency of the flexicle. However, no directed transport occurs. In contrast, rigid flexicles contain only latchers that push directly against the sphere with hardly any tangential motion on the object surface.  As a result, the sphere is steadily pushed in a linear trajectory with minimal relative motion between the flexicle and the object surface (see Fig.~\ref{fig:results1}(b) and Movie.2). To further explore the role of membrane deformability as a control parameter for transport, we dynamically alter the flexicle’s rigidity between $\kappa_H = 100~k_B\,T$ and $\kappa_H = 10000~k_B\,T$ over a time period of $150{,}000~\tau$. This modulation reversibly switches the system between oscillatory and persistent directional transport (see Movie.6 and Movie.7). These findings establish membrane bending rigidity as a key control parameter for actively tuning transport behavior.\\

Lastly, flexicles exhibit collective dynamics when multiple flexicles latch onto the same sphere. Relatively quickly, the flexicles orbiting motion becomes synchronized, (see Movie.8 with 2 flexicles and Movie.9 with 5 flexicles), suggesting that multiple flexicles can work together to enhance functionality.\\

  

\subsection{\label{sec:cylinder}Flexicle on a Cylinder}


\begin{figure*}[t!]
\adjincludegraphics[width = 0.9 \textwidth, trim={0} {.01\height} {.01\width} {.01\height},clip]{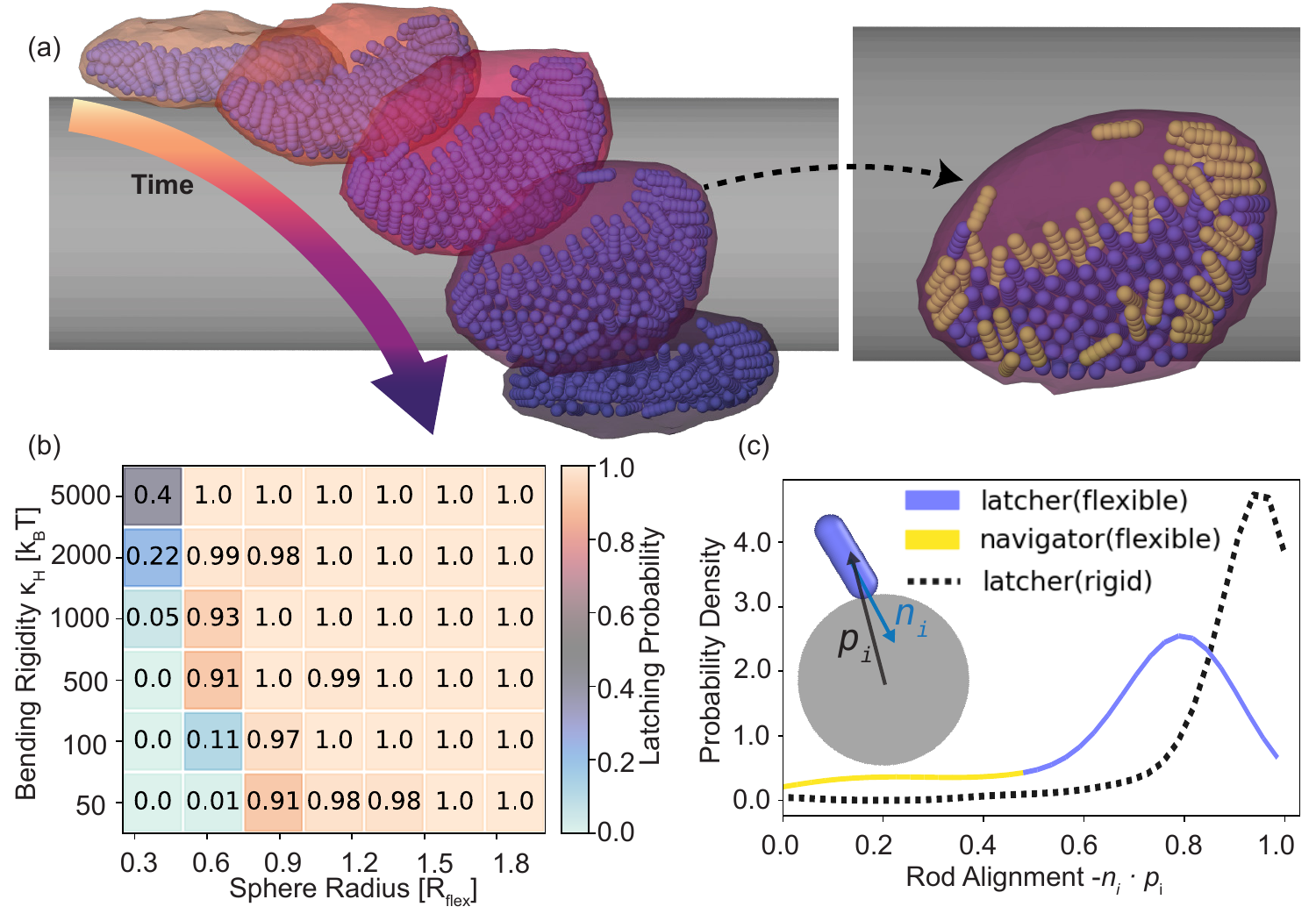}
\caption{\label{fig:results3} (a) Left: Snapshots showing a flexicle spiraling upward along a cylindrical surface. Right: Internal rod configuration, highlighting two functional roles—latchers (blue) and navigators (yellow). (b) Latching probability of flexicles on cylindrical obstacles. (c) Rod alignment distribution, comparing the rod's axis ($n_i$) with the local surface normal ($\hat{p}_i$). Solid and dotted lines represent flexible ($\kappa_H = 10~k_B\,T$) and rigid ($\kappa_H = 10000~k_B\,T$) membranes, respectively. Colored segments of the solid line correspond to latchers and navigators shown in (a). All simulations were performed at Péclet number $\mathrm{Pe} = 100$.}
\end{figure*}

When encountering cylindrical obstacles, flexicles exhibit a latching mechanism similar to that on spherical obstacles (see Movie.3). Upon collision, most internal rods align perpendicularly to the cylinder and press the flexicle against the obstacle. As in the case of sphere obstacles, the latching probability depends on the membrane bending rigidity. Flexicles with higher bending rigidity can attach to cylinders with smaller radii compared to their more deformable counterparts (see Fig.~\ref{fig:results3}(b)). We can again attribute this behavior to the difference in vesicle morphology when latched and the resulting larger effective radii for rigid flexicles (see Fig.~\ref{fig:results2}(b)). However, we find that the minimum cylinder radius required for stable latching is smaller than that needed for spherical obstacles. This reduction in the critical latching radius, as well as the increased stability of the contact between the flexicle and the cylinder, can be explained by the uniaxial curvature of the cylinder. Unlike a sphere, which curves in all directions, a cylinder is curved only along one axis and remains straight along its length. The same geometric argument applies to the detachment threshold. When gradually reducing the obstacle radius, the detachment radius $R_\text{detach}$ is systematically lower for cylinders than for spheres (see Fig.~\ref{fig:S4}).\\

Another shared feature between cylindrical and spherical obstacles is the spontaneous emergence of latchers and navigators. For rigid membranes with $\kappa_H = 10000~k_B\,T$, the distribution of rod orientation angles relative to the cylinder normal exhibits a single sharp peak, indicative of latchers. In contrast, more flexible membranes ($\kappa_H = 100~k_B\,T$) show a broader distribution with an additional peak at lower angles, corresponding to navigators (see Fig.~\ref{fig:results3}(c)). As on sphere surfaces, navigators dictate the direction of motion on the cylinder surface, driving the flexicles to orbit around the cylinder's symmetry axis (see Fig.~\ref{fig:results3}(a)). Because the cylinder perimeter is constant, small fluctuations result in spiral orbits, which can be right- or left-handed with equal probability. Fluctuations can also result in a sudden change of direction, turning right-handed motion along the cylinder axis into left-handed motion and vice versa.\\

\subsection{\label{sec:box}Flexicle in a Box and Climbing Stairs}

\begin{figure*}[t!]
  \includegraphics[width = 0.9 \textwidth]{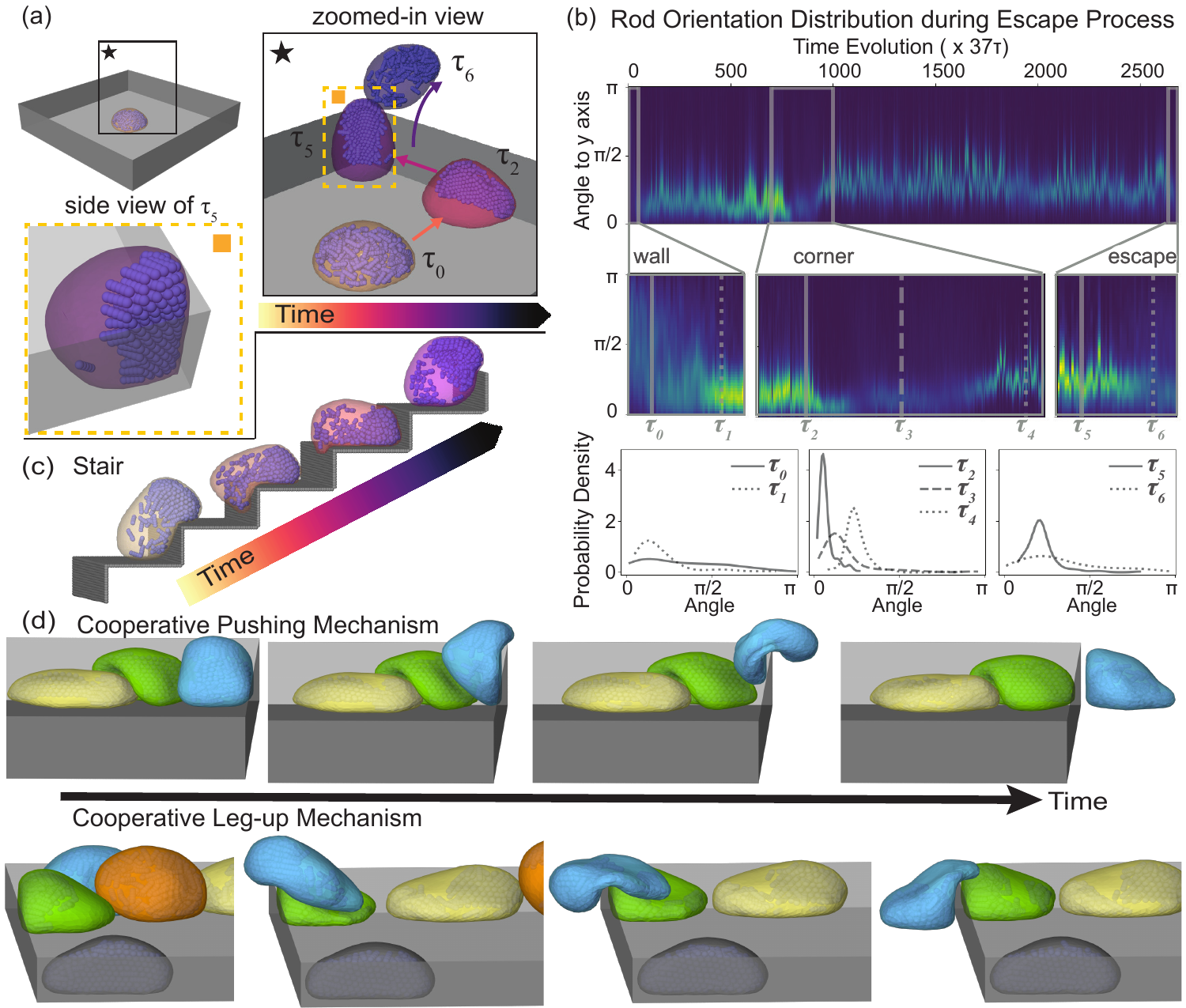}
  \caption{\label{fig:results4}(a) Snapshots showing a flexicle escaping from a square enclosure. Key time points: $\tau_0$ (center), $\tau_2$ (wall), $\tau_5$ (corner), and $\tau_6$ (escaped). Top-right inset shows a time-colored trajectory from a top view. Bottom-left inset shows rod alignment toward the corner at $\tau_5$. (b) Time evolution of rod orientation angles relative to the wall’s y-axis. Top: Full timeline. Middle: Zoom-ins on three key stages—(I) wall contact, (II) corner encounter, and (III) escape. Bottom: Angle distributions at specific times, comparing before, during and after each stage ($\tau_0$–$\tau_6$), with time points highlighted in the middle row. (c) Snapshots of a flexicle ($\kappa_H = 100~k_B\,T$) climbing a staircase (step height = $R_{\text{flex}}$) under a constant upward force $F_G = 3\,\epsilon/\sigma$. Color indicates progression in time. (d) Snapshots of two cooperative escape scenarios involving multiple flexicles. Each row shows a time series of four snapshots, with arrows indicating the progression over time. \textbf{Top row – Cooperative Pushing Mechanism}: Multiple flexicles work together to ``help" one flexicle (blue) climb a tall wall ($h_{\text{wall}} = 1.75~R_{\text{flex}}$). Yellow and green flexicles push from behind, compressing the blue flexicle. This compression enables internal rods in the blue flexicle to align upward and grow taller, allowing it to climb over the barrier. \textbf{Bottom row – Cooperative Leg-up Mechanism}: A group of flexicles collaborates to overcome a wall ($h_{\text{wall}} = 1.5~R_{\text{flex}}$). The yellow flexicle approaches the group, the orange flexicle exits, and the blue flexicle climbs over the wall by riding atop the green one. The rod alignment within the blue flexicle adjusts to support the climb.}
\end{figure*}

To further highlight the emergent microrobotic behaviors of flexicles that arise from the interplay between geometrical environment, vesicle shape, and internal cluster arrangement, we placed the flexicle inside a square box (see Fig.~\ref{fig:results4}(a)). In this scenario, the box is constructed with vertical walls of height $h$ and is open at the top. To prevent the flexicle from escaping by lifting vertically, we introduced an attractive interaction between the vertices of the flexicle mesh and the floor. This interaction mimics substrate adhesion, causing the flexicle to spread across the floor, increase its contact area, and adopt an oblate shape. Interaction with the walls is purely steric, with no attraction. Upon activation of the encapsulated self-propelled rods, the majority of the rods align parallel to the floor and function as navigators, propelling the flexicle across the enclosure. Eventually, the flexicle collides with one of the bounding walls. If $h\lesssim6~\sigma\simeq0.75~R_\text{flex}$ is smaller than the height of the internal rod cluster, the cluster moves the entire flexicle over the wall with hardly any cluster rearrangements. However, when the wall height exceeds this threshold, the collision significantly perturbs the rod alignment and disrupts the forward motion. In that case, a subset of rods reorients nearly perpendicularly to the boundary, forming a planar wall-facing configuration. These rods act as latchers, pressing the flexicle against the wall and stabilizing it. This first transition from predominantly navigators to predominantly latchers can be interpreted as an adaptive response to confinement and marks the beginning of a multistage escape process.\\

Once a stable interaction with the wall is established, the flexicle begins to drift along the boundary, driven by a small fraction of rods that remain navigators. These navigators induce an imbalance in the cluster’s propulsion direction, steering the flexicle toward a corner. At the corner where two walls meet at a right angle, the corner geometry promotes a more compact rod configuration. This arises from the membrane conforming to the corner and spontaneously forming a sharply curved, kink-like domain. Previous studies on 2D flexicle-like particle systems with pre-designed kinks in their membranes demonstrated that such geometrical discontinuities act as focal points for rod clustering and active reorganization~\cite{lee_complex_2023}. Similarly, here the rods reorient and accumulate at the corner, leading to an increase in cluster height. The cluster growth is accompanied by further dynamic role-switching within the cluster: latchers at the corner exert sustained pressure on the boundary, while navigators at the top of the cluster push upwards and generate bulges. This deformation increases the vertical height of the flexicle and enables upward movement. If the growing cluster extends above the wall height, the top-facing latchers no longer encounter external resistance and instead act as navigators, pushing the flexicle over the boundary. This upward propulsion ultimately allows the flexicle to escape the enclosure, even when the wall height exceeds the initial flexicle size (see Movie.4). In simulations with $\kappa_H=100\,k_B\,T$ and $N_\text{rod}=153$ flexicles successfully overcome walls as tall as $h_\text{max}\simeq18~\sigma\simeq 2.25~R_\text{flex}$. Surrounded by higher walls, and for the range of flexicle parameters studied, flexicles fail to generate sufficient vertical extension and remain trapped in the corner. \\

As we have seen, the entire escape sequence is governed by a dynamic redistribution of roles among the internal self-propelled rods, driven by their relative orientation and local spatial constraints at any given time. This process is reflected in the distribution of rod orientations along the y-axis during a typical escape event, as shown in Fig.~\ref{fig:results4}(b). In stage I, while navigating the interior of the box, the angle distribution is broad, indicating a disordered configuration with rods pointing in many directions. In stage II, while the flexicle glides along the wall, the angle distribution narrows to a sharp peak at $\alpha = 0$, corresponding to a highly ordered alignment parallel to the wall. In stage III, as the flexicle grows vertically at the corner, the distribution shifts toward $\alpha = \pi/4$, consistent with the formation of a tilted cluster oriented along the corner’s bisector. Finally, in stage IV, as the flexicle climbs and exits the enclosure, the angular distribution becomes diffuse once again, indicating the escape. This transition between internal organization states underscores the flexicle's capacity to adapt to changing environments. Without external programming or centralized control, the flexicle leverages local rod-membrane interactions and environmental geometry to generate complex, functional behavior, transforming a static confinement scenario into a pathway for escape. This emergent behavior almost makes it appear as though the flexicles have agency, which they do not.\\

The adaptive escape mechanism described above is not limited to flexicles that adhere to a substrate through attractive interactions, but also applies to flexicles subjected to a constant force, such as gravity, that keeps them in contact with the floor. By applying a constant force $F_G = 3\,\epsilon/\sigma$ to each mesh vertex in the $z$-direction, we can reproduce qualitatively similar escape trajectories, including the entire four-stage escape sequence and the characteristic reorganizations of the internal rod cluster (see Fig.~\ref{fig:S8}). However, the maximum wall height $h_\text{max}\simeq12~\sigma\simeq 1.5~R_\text{flex}$ that these flexicles can overcome is lower compared to flexicles that adhere solely through surface adhesion. Presumably, higher walls could be surmounted by adding an attraction between the membrane and the wall, in addition to the floor.  

Nevertheless, flexicles attracted in some way to a substrate benefit from being less prone to detachment and uncontrolled motion after escape, even upon temporarily losing contact with the floor. This benefit allows them to repeatedly overcome a consecutive series of wall obstacles, such as is needed to climb a set of stairs (see Fig.~\ref{fig:results4}(c) and Movie.5).\\

Extending the concept of environment-affected behavior, we find that groups of flexicles are capable of cooperative interactions that enable them to overcome obstacles impassable to individual flexicles. In one simulation, for example, multiple flexicles accumulate at a common corner, creating a crowded configuration in which the leading flexicle, closest to the corner, experiences compressive forces from those behind. This pressure causes an additional vertical deformation of the leading flexicle, which in turn triggers a realignment of its internal rods toward its upward axis. The resulting reorganization generates sufficient propulsion to lift the leading flexicle over a wall that would otherwise exceed its individual climbing capability (see Fig.\ref{fig:results4}(d) and Movie.10). In an alternative cooperative scenario, we observe that the lead flexicle can instead act as a physical ``step up". Here, the following flexicles climb over the lead flexicle, effectively using it as a temporary staircase (see Fig.\ref{fig:results4}(d) and Movie.11).\\

\section{\label{sec:discussion}Discussion and Conclusions}

In this study, we investigated the emergent behaviors of active, deformable compound particles or flexicles composed of self-propelled colloidal rods enclosed within vesicles. When encountering various external obstacles, flexicles exhibited a range of dynamic behaviors, including latching onto curved surfaces, gliding along interfaces, navigating toward geometric features such as corners, and even climbing over barriers to escape confinement or traverse a staircase. These phenomena arise from the complex interplay between vesicle deformation, internal rod arrangement, and the geometry of the environment. Our simulations demonstrate that flexicles offer a platform for designing microparticle systems capable of executing different robotic tasks. All of the observed geometry-dependent dynamic behaviors are based on emergent functionality rather than explicitly pre-programmed functions, highlighting their potential as small-scale robotic agents.\\

We found that these microrobotic behaviors are primarily governed by the spontaneous rearrangement of the self-propelled internal rod cluster in response to the flexicle's collision with the external object. Upon contact with an obstacle, a significant portion of the internal rods, which we term latchers, reorient perpendicular to the surface. Latchers collectively exert forces against the inner membrane, enabling the flexicle to attach to curved objects such as spheres and cylinders. Moreover, we identified a minimum curvature threshold required for stable attachment, which depends on membrane rigidity. Additionally, the interactions between the flexicle and the object can induce significant vesicle deformations. For flexicles with low bending rigidity, the membrane conforms closely to the obstacle's contour. This morphological adaptability promotes the emergence of a secondary population of rods that align mostly tangentially to the surface. We term these rods navigators, as their propulsion direction introduces an asymmetry in the net force balance, causing the flexicle to slide along the obstacle's surface. This behavior marks a rigidity-controlled transition from statically latched, rigid flexicles to dynamically orbiting or spiraling deformable flexicles. Moreover, we demonstrated that this combination of latching and guided surface motion can be exploited to transport movable objects.\\

We demonstrated that the dynamic exchange between latchers and navigators enables flexicles to escape from confined environments, such as square boxes. Through continuous reorganization of their internal rod cluster, flexicles can spontaneously locate and orient along geometric features like walls and corners. This adaptive redistribution allows them to transition from horizontal motion to vertical climbing, ultimately surmounting barriers significantly taller than their own size without any form of external control or programming.\\

Our simulations of flexicles with dynamic membrane rigidity suggest that introducing time-dependent mechanical features can give rise to even more versatile and adaptive robotic behaviors~\cite{veenstra_adaptive_2025}. Biological cells, for instance, fluidize or stiffen their membranes through processes like polymerization to modulate locomotion and their environmental response~\cite{greene_role_2016}. Drawing inspiration from these strategies observed in nature, flexicles equipped with tunable or responsive membrane properties could exhibit advanced behaviors such as selective trapping, directional release, or mode switching—analogous to catch-and-release mechanisms seen in biological systems~\cite{sarangapani_catch_2014}. Our demonstrations of rigidity-dependent dynamics underscore membrane bending rigidity as a key design parameter for developing flexicle-based microrobots with task-specific adaptability.\\

Another promising direction for future research lies in the exploration of collective robotic behaviors in vesicle systems. Both biological organisms and engineered microrobots often rely on cooperation among multiple agents to perform tasks that exceed the capabilities of individual units. For instance, the bacterium \textit{Myxococcus xanthus} exhibits cooperative swarming behavior, forming coordinated "wolf-pack" structures to prey on other microbes more effectively~\cite{berleman_deciphering_2009}. Likewise, social animals like ant colonies demonstrate decentralized coordination, enabling them to construct intricate nests and transport large food items collectively~\cite{gordon_interaction_2010}. Recent developments in microrobotics have similarly demonstrated that swarms of micro- and nanoscale robots can transport cargo collaboratively by leveraging local interactions and distributed control strategies~\cite{gardi_microrobot_2022}. Inspired by these systems, vesicle-based collectives may offer new microrobotic strategies for traversing complex environments or performing group-level functions. Our findings suggest that inter-vesicle interactions, such as mechanical contact or local reinforcement, can facilitate cooperative dynamics. Future work should more systematically investigate how such interactions depend on design parameters, including membrane stiffness, vesicle density, or shape, to enable programmable teamwork, scaffold formation, or cargo manipulation in confined or unstructured settings. By advancing this line of research, vesicle systems could serve as a biomimetic platform for exploring collective intelligence and adaptive function in soft microrobotics.\\

Although our simulations offer valuable theoretical insights, experimental validation will be crucial to fully unlocking the practical potential of vesicle-based systems. In particular, the ability to precisely control membrane stiffness is key to tuning the mechanical behavior of vesicles, synthetic cells, and biomimetic membranes. Most membranes that can be synthesized today fall within the lower range of bending rigidities we explored. For example, lipid bilayers typically exhibit bending rigidities up to ~$150~k_B\,T$, which can be modulated by adjusting lipid saturation~\cite{tarazona_thermal_2013}, chain length~\cite{rawicz_effect_2000}, membrane thickness~\cite{bermudez_effect_2004}, or by incorporating cholesterol~\cite{doole_cholesterol_2022}, charged lipids~\cite{faizi_bending_2019}, or polymeric/solid supports~\cite{fowler_membrane_2016,lopresti_polymersomes_2009, lasic_sterically_1994}. Polymersome-based vesicles, formed from amphiphilic block copolymers, represent another widely used system and often exhibit higher bending rigidity and enhanced mechanical stability, making them suitable candidates for realizing flexicles with high bending rigidities~\cite{faizi_bending_2022, bermudez_effect_2004, rideau_liposomes_2018, lopresti_polymersomes_2009, discher_polymersomes_1999}.\\

Looking ahead, future work should focus on developing strategies for dynamic stiffness control, optimizing membrane design for task-specific performance, and exploring hybrid systems that integrate biological components to enhance the functional complexity of vesicle-based microrobots. Furthermore, the influence of hydrodynamic interactions and fluid-mediated effects deserves deeper investigation to more accurately reflect realistic operating environments.\\

In summary, this study provides foundational insight into the design and control of vesicle-encapsulated active matter systems. By emphasizing the roles of rod alignment and membrane flexibility, our findings lay the groundwork for developing sophisticated, adaptable microrobotic systems that combine aspects of colloidal and biological physics.

\begin{acknowledgments}
This work was supported by a Vannevar Bush Faculty Fellowship sponsored by the Department of the Navy, Office of Naval Research under ONR award number N00014-22-1-2821. Computational work was carried out on Anvil CPU at Purdue, Delta GPU at NCSA, and Bridges2 CPU at PSC through allocation DMR 140129 from the Advanced Cyberinfrastructure Coordination Ecosystem: Services \& Support (ACCESS) program, which is supported by National Science Foundation grants \#2138259, \#2138286, \#2138307, \#2137603, and \#2138296. Computational resources and services were also provided by Advanced Research Computing at the University of Michigan, Ann Arbor.
\end{acknowledgments}





\bibliography{references}

\begin{thebibliography}{89}
\providecommand{\natexlab}[1]{#1}
\providecommand{\url}[1]{\texttt{#1}}
\expandafter\ifx\csname urlstyle\endcsname\relax
  \providecommand{\doi}[1]{doi: #1}\else
  \providecommand{\doi}{doi: \begingroup \urlstyle{rm}\Url}\fi

\bibitem[Liu et~al.(2023)Liu, Hempel, Yang, Brooks, Pervan, Koman, Zhang, Kozawa, Yang, Goldman, Miskin, Richa, Randall, Murphey, Palacios, and Strano]{liu_colloidal_2023}
Albert~Tianxiang Liu, Marek Hempel, Jing~Fan Yang, Allan~M. Brooks, Ana Pervan, Volodymyr~B. Koman, Ge~Zhang, Daichi Kozawa, Sungyun Yang, Daniel~I. Goldman, Marc~Z. Miskin, Andréa~W. Richa, Dana Randall, Todd~D. Murphey, Tomás Palacios, and Michael~S. Strano.
\newblock Colloidal robotics.
\newblock \emph{Nature Materials}, 22\penalty0 (12):\penalty0 1453--1462, December 2023.
\newblock ISSN 1476-1122, 1476-4660.
\newblock \doi{10.1038/s41563-023-01589-y}.
\newblock URL \url{https://www.nature.com/articles/s41563-023-01589-y}.

\bibitem[Bishop et~al.(2023)Bishop, Biswal, and Bharti]{bishop_active_2023}
Kyle~J.M. Bishop, Sibani~Lisa Biswal, and Bhuvnesh Bharti.
\newblock Active {Colloids} as {Models}, {Materials}, and {Machines}.
\newblock \emph{Annual Review of Chemical and Biomolecular Engineering}, 14\penalty0 (1):\penalty0 1--30, June 2023.
\newblock ISSN 1947-5438, 1947-5446.
\newblock \doi{10.1146/annurev-chembioeng-101121-084939}.
\newblock URL \url{https://www.annualreviews.org/doi/10.1146/annurev-chembioeng-101121-084939}.

\bibitem[Vicsek et~al.(1995)Vicsek, Czirók, Ben-Jacob, Cohen, and Shochet]{vicsek_novel_1995}
Tamás Vicsek, András Czirók, Eshel Ben-Jacob, Inon Cohen, and Ofer Shochet.
\newblock Novel {Type} of {Phase} {Transition} in a {System} of {Self}-{Driven} {Particles}.
\newblock \emph{Physical Review Letters}, 75\penalty0 (6):\penalty0 1226--1229, August 1995.
\newblock ISSN 0031-9007, 1079-7114.
\newblock \doi{10.1103/PhysRevLett.75.1226}.
\newblock URL \url{https://link.aps.org/doi/10.1103/PhysRevLett.75.1226}.

\bibitem[Ballerini et~al.(2008)Ballerini, Cabibbo, Candelier, Cavagna, Cisbani, Giardina, Lecomte, Orlandi, Parisi, Procaccini, Viale, and Zdravkovic]{ballerini_interaction_2008}
M.~Ballerini, N.~Cabibbo, R.~Candelier, A.~Cavagna, E.~Cisbani, I.~Giardina, V.~Lecomte, A.~Orlandi, G.~Parisi, A.~Procaccini, M.~Viale, and V.~Zdravkovic.
\newblock Interaction ruling animal collective behavior depends on topological rather than metric distance: {Evidence} from a field study.
\newblock \emph{Proceedings of the National Academy of Sciences}, 105\penalty0 (4):\penalty0 1232--1237, January 2008.
\newblock ISSN 0027-8424, 1091-6490.
\newblock \doi{10.1073/pnas.0711437105}.
\newblock URL \url{https://pnas.org/doi/full/10.1073/pnas.0711437105}.

\bibitem[Schaller et~al.(2010)Schaller, Weber, Semmrich, Frey, and Bausch]{schaller_polar_2010}
Volker Schaller, Christoph Weber, Christine Semmrich, Erwin Frey, and Andreas~R. Bausch.
\newblock Polar patterns of driven filaments.
\newblock \emph{Nature}, 467\penalty0 (7311):\penalty0 73--77, September 2010.
\newblock ISSN 0028-0836, 1476-4687.
\newblock \doi{10.1038/nature09312}.
\newblock URL \url{https://www.nature.com/articles/nature09312}.

\bibitem[Palacci et~al.(2013)Palacci, Sacanna, Steinberg, Pine, and Chaikin]{palacci_living_2013}
Jeremie Palacci, Stefano Sacanna, Asher~Preska Steinberg, David~J. Pine, and Paul~M. Chaikin.
\newblock Living {Crystals} of {Light}-{Activated} {Colloidal} {Surfers}.
\newblock \emph{Science}, 339\penalty0 (6122):\penalty0 936--940, February 2013.
\newblock ISSN 0036-8075, 1095-9203.
\newblock \doi{10.1126/science.1230020}.
\newblock URL \url{https://www.science.org/doi/10.1126/science.1230020}.

\bibitem[Deseigne et~al.(2012)Deseigne, Léonard, Dauchot, and Chaté]{deseigne_vibrated_2012}
Julien Deseigne, Sébastien Léonard, Olivier Dauchot, and Hugues Chaté.
\newblock Vibrated polar disks: spontaneous motion, binary collisions, and collective dynamics.
\newblock \emph{Soft Matter}, 8\penalty0 (20):\penalty0 5629, 2012.
\newblock ISSN 1744-683X, 1744-6848.
\newblock \doi{10.1039/c2sm25186h}.
\newblock URL \url{https://xlink.rsc.org/?DOI=c2sm25186h}.

\bibitem[Gumuskaya et~al.(2023)Gumuskaya, Srivastava, Cooper, Lesser, Semegran, Garnier, and Levin]{gumuskaya_motile_2023}
Gizem Gumuskaya, Pranjal Srivastava, Ben~G. Cooper, Hannah Lesser, Ben Semegran, Simon Garnier, and Michael Levin.
\newblock Motile {Living} {Biobots} {Self}‐{Construct} from {Adult} {Human} {Somatic} {Progenitor} {Seed} {Cells}.
\newblock \emph{Advanced Science}, page 2303575, November 2023.
\newblock ISSN 2198-3844, 2198-3844.
\newblock \doi{10.1002/advs.202303575}.
\newblock URL \url{https://onlinelibrary.wiley.com/doi/10.1002/advs.202303575}.

\bibitem[Park et~al.(2016)Park, Kim, Kim, Wang, Kwak, Hur, Suh, An, and Levchenko]{park_directed_2016}
JinSeok Park, Deok-Ho Kim, Hong-Nam Kim, Chiaochun~Joanne Wang, Moon~Kyu Kwak, Eunmi Hur, Kahp-Yang Suh, Steven~S. An, and Andre Levchenko.
\newblock Directed migration of cancer cells guided by the graded texture of the underlying matrix.
\newblock \emph{Nature Materials}, 15\penalty0 (7):\penalty0 792--801, July 2016.
\newblock ISSN 1476-1122, 1476-4660.
\newblock \doi{10.1038/nmat4586}.
\newblock URL \url{https://www.nature.com/articles/nmat4586}.

\bibitem[Lushi et~al.(2012)Lushi, Goldstein, and Shelley]{lushi_collective_2012}
Enkeleida Lushi, Raymond~E. Goldstein, and Michael~J. Shelley.
\newblock Collective chemotactic dynamics in the presence of self-generated fluid flows.
\newblock \emph{Physical Review E}, 86\penalty0 (4):\penalty0 040902, October 2012.
\newblock ISSN 1539-3755, 1550-2376.
\newblock \doi{10.1103/PhysRevE.86.040902}.
\newblock URL \url{https://link.aps.org/doi/10.1103/PhysRevE.86.040902}.

\bibitem[Reversat et~al.(2020)Reversat, Gaertner, Merrin, Stopp, Tasciyan, Aguilera, De~Vries, Hauschild, Hons, Piel, Callan-Jones, Voituriez, and Sixt]{reversat_cellular_2020}
Anne Reversat, Florian Gaertner, Jack Merrin, Julian Stopp, Saren Tasciyan, Juan Aguilera, Ingrid De~Vries, Robert Hauschild, Miroslav Hons, Matthieu Piel, Andrew Callan-Jones, Raphael Voituriez, and Michael Sixt.
\newblock Cellular locomotion using environmental topography.
\newblock \emph{Nature}, 582\penalty0 (7813):\penalty0 582--585, June 2020.
\newblock ISSN 0028-0836, 1476-4687.
\newblock \doi{10.1038/s41586-020-2283-z}.
\newblock URL \url{https://www.nature.com/articles/s41586-020-2283-z}.

\bibitem[Keya et~al.(2018)Keya, Suzuki, Kabir, Inoue, Asanuma, Sada, Hess, Kuzuya, and Kakugo]{keya_dna-assisted_2018}
Jakia~Jannat Keya, Ryuhei Suzuki, Arif Md.~Rashedul Kabir, Daisuke Inoue, Hiroyuki Asanuma, Kazuki Sada, Henry Hess, Akinori Kuzuya, and Akira Kakugo.
\newblock {DNA}-assisted swarm control in a biomolecular motor system.
\newblock \emph{Nature Communications}, 9\penalty0 (1):\penalty0 453, January 2018.
\newblock ISSN 2041-1723.
\newblock \doi{10.1038/s41467-017-02778-5}.
\newblock URL \url{https://www.nature.com/articles/s41467-017-02778-5}.

\bibitem[Li et~al.(2018)Li, Li, Luo, Wang, Liu, Chen, Li, Yue, Cheng, and Sun]{li_development_2018}
Junyang Li, Xiaojian Li, Tao Luo, Ran Wang, Chichi Liu, Shuxun Chen, Dongfang Li, Jianbo Yue, Shuk-han Cheng, and D.~Sun.
\newblock Development of a magnetic microrobot for carrying and delivering targeted cells.
\newblock \emph{Science Robotics}, 3\penalty0 (19):\penalty0 eaat8829, June 2018.
\newblock ISSN 2470-9476.
\newblock \doi{10.1126/scirobotics.aat8829}.
\newblock URL \url{https://www.science.org/doi/10.1126/scirobotics.aat8829}.

\bibitem[Solovev et~al.(2010)Solovev, Sanchez, Pumera, Mei, and Schmidt]{solovev_magnetic_2010}
Alexander~A. Solovev, Samuel Sanchez, Martin Pumera, Yong~Feng Mei, and Oliver~G. Schmidt.
\newblock Magnetic {Control} of {Tubular} {Catalytic} {Microbots} for the {Transport}, {Assembly}, and {Delivery} of {Micro}‐objects.
\newblock \emph{Advanced Functional Materials}, 20\penalty0 (15):\penalty0 2430--2435, August 2010.
\newblock ISSN 1616-301X, 1616-3028.
\newblock \doi{10.1002/adfm.200902376}.
\newblock URL \url{https://onlinelibrary.wiley.com/doi/10.1002/adfm.200902376}.

\bibitem[Nelson and Peyer(2014)]{nelson_micro-_2014}
Bradley~J. Nelson and Kathrin~E. Peyer.
\newblock Micro- and {Nanorobots} {Swimming} in {Heterogeneous} {Liquids}.
\newblock \emph{ACS Nano}, 8\penalty0 (9):\penalty0 8718--8724, September 2014.
\newblock ISSN 1936-0851, 1936-086X.
\newblock \doi{10.1021/nn504295z}.
\newblock URL \url{https://pubs.acs.org/doi/10.1021/nn504295z}.

\bibitem[Gardi et~al.(2022)Gardi, Ceron, Wang, Petersen, and Sitti]{gardi_microrobot_2022}
Gaurav Gardi, Steven Ceron, Wendong Wang, Kirstin Petersen, and Metin Sitti.
\newblock Microrobot collectives with reconfigurable morphologies, behaviors, and functions.
\newblock \emph{Nature Communications}, 13\penalty0 (1):\penalty0 2239, April 2022.
\newblock ISSN 2041-1723.
\newblock \doi{10.1038/s41467-022-29882-5}.
\newblock URL \url{https://www.nature.com/articles/s41467-022-29882-5}.

\bibitem[Akter et~al.(2022)Akter, Keya, Kayano, Kabir, Inoue, Hess, Sada, Kuzuya, Asanuma, and Kakugo]{akter_cooperative_2022}
M.~Akter, J.~J. Keya, K.~Kayano, A.~M.~R. Kabir, D.~Inoue, H.~Hess, K.~Sada, A.~Kuzuya, H.~Asanuma, and A.~Kakugo.
\newblock Cooperative cargo transportation by a swarm of molecular machines.
\newblock \emph{Science Robotics}, 7\penalty0 (65):\penalty0 eabm0677, April 2022.
\newblock ISSN 2470-9476.
\newblock \doi{10.1126/scirobotics.abm0677}.
\newblock URL \url{https://www.science.org/doi/10.1126/scirobotics.abm0677}.

\bibitem[Zhang et~al.(2009)Zhang, Abbott, Dong, Kratochvil, Bell, and Nelson]{zhang_artificial_2009}
Li~Zhang, Jake~J. Abbott, Lixin Dong, Bradley~E. Kratochvil, Dominik Bell, and Bradley~J. Nelson.
\newblock Artificial bacterial flagella: {Fabrication} and magnetic control.
\newblock \emph{Applied Physics Letters}, 94\penalty0 (6):\penalty0 064107, February 2009.
\newblock ISSN 0003-6951, 1077-3118.
\newblock \doi{10.1063/1.3079655}.
\newblock URL \url{https://pubs.aip.org/apl/article/94/6/064107/120988/Artificial-bacterial-flagella-Fabrication-and}.

\bibitem[Ghosh and Fischer(2009)]{ghosh_controlled_2009}
Ambarish Ghosh and Peer Fischer.
\newblock Controlled {Propulsion} of {Artificial} {Magnetic} {Nanostructured} {Propellers}.
\newblock \emph{Nano Letters}, 9\penalty0 (6):\penalty0 2243--2245, June 2009.
\newblock ISSN 1530-6984, 1530-6992.
\newblock \doi{10.1021/nl900186w}.
\newblock URL \url{https://pubs.acs.org/doi/10.1021/nl900186w}.

\bibitem[Tierno et~al.(2008)Tierno, Golestanian, Pagonabarraga, and Sagués]{tierno_controlled_2008}
Pietro Tierno, Ramin Golestanian, Ignacio Pagonabarraga, and Francesc Sagués.
\newblock Controlled {Swimming} in {Confined} {Fluids} of {Magnetically} {Actuated} {Colloidal} {Rotors}.
\newblock \emph{Physical Review Letters}, 101\penalty0 (21):\penalty0 218304, November 2008.
\newblock ISSN 0031-9007, 1079-7114.
\newblock \doi{10.1103/PhysRevLett.101.218304}.
\newblock URL \url{https://link.aps.org/doi/10.1103/PhysRevLett.101.218304}.

\bibitem[Akolpoglu et~al.(2022)Akolpoglu, Alapan, Dogan, Baltaci, Yasa, Aybar~Tural, and Sitti]{akolpoglu_magnetically_2022}
Mukrime~Birgul Akolpoglu, Yunus Alapan, Nihal~Olcay Dogan, Saadet~Fatma Baltaci, Oncay Yasa, Gulsen Aybar~Tural, and Metin Sitti.
\newblock Magnetically steerable bacterial microrobots moving in {3D} biological matrices for stimuli-responsive cargo delivery.
\newblock \emph{Science Advances}, 8\penalty0 (28):\penalty0 eabo6163, July 2022.
\newblock ISSN 2375-2548.
\newblock \doi{10.1126/sciadv.abo6163}.
\newblock URL \url{https://www.science.org/doi/10.1126/sciadv.abo6163}.

\bibitem[Huang et~al.(2016)Huang, Sakar, Petruska, Pané, and Nelson]{huang_soft_2016}
Hen-Wei Huang, Mahmut~Selman Sakar, Andrew~J. Petruska, Salvador Pané, and Bradley~J. Nelson.
\newblock Soft micromachines with programmable motility and morphology.
\newblock \emph{Nature Communications}, 7\penalty0 (1):\penalty0 12263, July 2016.
\newblock ISSN 2041-1723.
\newblock \doi{10.1038/ncomms12263}.
\newblock URL \url{https://www.nature.com/articles/ncomms12263}.

\bibitem[Jiang et~al.(2010)Jiang, Yoshinaga, and Sano]{jiang_active_2010}
Hong-Ren Jiang, Natsuhiko Yoshinaga, and Masaki Sano.
\newblock Active {Motion} of a {Janus} {Particle} by {Self}-{Thermophoresis} in a {Defocused} {Laser} {Beam}.
\newblock \emph{Physical Review Letters}, 105\penalty0 (26):\penalty0 268302, December 2010.
\newblock ISSN 0031-9007, 1079-7114.
\newblock \doi{10.1103/PhysRevLett.105.268302}.
\newblock URL \url{https://link.aps.org/doi/10.1103/PhysRevLett.105.268302}.

\bibitem[Debata et~al.(2024)Debata, Sahu, Panda, and Singh]{debata_light-driven_2024}
Srikanta Debata, Shivani Sahu, Suvendu~Kumar Panda, and Dhruv~Pratap Singh.
\newblock Light-driven micromotors for on-demand and local {pH} sensing applications.
\newblock \emph{Journal of Materials Chemistry B}, 12\penalty0 (8):\penalty0 2150--2157, 2024.
\newblock ISSN 2050-750X, 2050-7518.
\newblock \doi{10.1039/D3TB02760K}.
\newblock URL \url{https://xlink.rsc.org/?DOI=D3TB02760K}.

\bibitem[Palagi et~al.(2019)Palagi, Singh, and Fischer]{palagi_lightcontrolled_2019}
Stefano Palagi, Dhruv~P. Singh, and Peer Fischer.
\newblock Light‐{Controlled} {Micromotors} and {Soft} {Microrobots}.
\newblock \emph{Advanced Optical Materials}, 7\penalty0 (16):\penalty0 1900370, August 2019.
\newblock ISSN 2195-1071, 2195-1071.
\newblock \doi{10.1002/adom.201900370}.
\newblock URL \url{https://onlinelibrary.wiley.com/doi/10.1002/adom.201900370}.

\bibitem[Collins et~al.(2016)Collins, Devendran, Ma, Ng, Neild, and Ai]{collins_acoustic_2016}
David~J. Collins, Citsabehsan Devendran, Zhichao Ma, Jia~Wei Ng, Adrian Neild, and Ye~Ai.
\newblock Acoustic tweezers via sub–time-of-flight regime surface acoustic waves.
\newblock \emph{Science Advances}, 2\penalty0 (7):\penalty0 e1600089, July 2016.
\newblock ISSN 2375-2548.
\newblock \doi{10.1126/sciadv.1600089}.
\newblock URL \url{https://www.science.org/doi/10.1126/sciadv.1600089}.

\bibitem[Ahmed et~al.(2016)Ahmed, Baasch, Jang, Pane, Dual, and Nelson]{ahmed_artificial_2016}
Daniel Ahmed, Thierry Baasch, Bumjin Jang, Salvador Pane, Jürg Dual, and Bradley~J. Nelson.
\newblock Artificial {Swimmers} {Propelled} by {Acoustically} {Activated} {Flagella}.
\newblock \emph{Nano Letters}, 16\penalty0 (8):\penalty0 4968--4974, August 2016.
\newblock ISSN 1530-6984, 1530-6992.
\newblock \doi{10.1021/acs.nanolett.6b01601}.
\newblock URL \url{https://pubs.acs.org/doi/10.1021/acs.nanolett.6b01601}.

\bibitem[Wang et~al.(2012)Wang, Castro, Hoyos, and Mallouk]{wang_autonomous_2012}
Wei Wang, Luz~Angelica Castro, Mauricio Hoyos, and Thomas~E. Mallouk.
\newblock Autonomous {Motion} of {Metallic} {Microrods} {Propelled} by {Ultrasound}.
\newblock \emph{ACS Nano}, 6\penalty0 (7):\penalty0 6122--6132, July 2012.
\newblock ISSN 1936-0851, 1936-086X.
\newblock \doi{10.1021/nn301312z}.
\newblock URL \url{https://pubs.acs.org/doi/10.1021/nn301312z}.

\bibitem[Qiu et~al.(2014)Qiu, Lee, Mark, Morozov, Münster, Mierka, Turek, Leshansky, and Fischer]{qiu_swimming_2014}
Tian Qiu, Tung-Chun Lee, Andrew~G. Mark, Konstantin~I. Morozov, Raphael Münster, Otto Mierka, Stefan Turek, Alexander~M. Leshansky, and Peer Fischer.
\newblock Swimming by reciprocal motion at low {Reynolds} number.
\newblock \emph{Nature Communications}, 5\penalty0 (1):\penalty0 5119, November 2014.
\newblock ISSN 2041-1723.
\newblock \doi{10.1038/ncomms6119}.
\newblock URL \url{https://www.nature.com/articles/ncomms6119}.

\bibitem[Garcia‐Gradilla et~al.(2014)Garcia‐Gradilla, Sattayasamitsathit, Soto, Kuralay, Yardımcı, Wiitala, Galarnyk, and Wang]{garciagradilla_ultrasoundpropelled_2014}
Victor Garcia‐Gradilla, Sirilak Sattayasamitsathit, Fernando Soto, Filiz Kuralay, Ceren Yardımcı, Devan Wiitala, Michael Galarnyk, and Joseph Wang.
\newblock Ultrasound‐{Propelled} {Nanoporous} {Gold} {Wire} for {Efficient} {Drug} {Loading} and {Release}.
\newblock \emph{Small}, 10\penalty0 (20):\penalty0 4154--4159, October 2014.
\newblock ISSN 1613-6810, 1613-6829.
\newblock \doi{10.1002/smll.201401013}.
\newblock URL \url{https://onlinelibrary.wiley.com/doi/10.1002/smll.201401013}.

\bibitem[Dong et~al.(2016)Dong, Zhang, Gao, Pei, and Ren]{dong_highly_2016}
Renfeng Dong, Qilu Zhang, Wei Gao, Allen Pei, and Biye Ren.
\newblock Highly {Efficient} {Light}-{Driven} {TiO}$_{\textrm{2}}$ –{Au} {Janus} {Micromotors}.
\newblock \emph{ACS Nano}, 10\penalty0 (1):\penalty0 839--844, January 2016.
\newblock ISSN 1936-0851, 1936-086X.
\newblock \doi{10.1021/acsnano.5b05940}.
\newblock URL \url{https://pubs.acs.org/doi/10.1021/acsnano.5b05940}.

\bibitem[Li et~al.(2016)Li, Li, Zhang, Chang, Song, Hu, Shao, Sandraz, Zhang, Li, and Wang]{li_magnetically_2016}
Tianlong Li, Jinxing Li, Hongtao Zhang, Xiaocong Chang, Wenping Song, Yanan Hu, Guangbin Shao, Elodie Sandraz, Guangyu Zhang, Longqiu Li, and Joseph Wang.
\newblock Magnetically {Propelled} {Fish}‐{Like} {Nanoswimmers}.
\newblock \emph{Small}, 12\penalty0 (44):\penalty0 6098--6105, November 2016.
\newblock ISSN 1613-6810, 1613-6829.
\newblock \doi{10.1002/smll.201601846}.
\newblock URL \url{https://onlinelibrary.wiley.com/doi/10.1002/smll.201601846}.

\bibitem[Demirörs et~al.(2018)Demirörs, Akan, Poloni, and Studart]{demirors_active_2018}
Ahmet~F. Demirörs, Mehmet~Tolga Akan, Erik Poloni, and André~R. Studart.
\newblock Active cargo transport with {Janus} colloidal shuttles using electric and magnetic fields.
\newblock \emph{Soft Matter}, 14\penalty0 (23):\penalty0 4741--4749, 2018.
\newblock ISSN 1744-683X, 1744-6848.
\newblock \doi{10.1039/C8SM00513C}.
\newblock URL \url{https://xlink.rsc.org/?DOI=C8SM00513C}.

\bibitem[Lum et~al.(2016)Lum, Ye, Dong, Marvi, Erin, Hu, and Sitti]{lum_shape-programmable_2016}
Guo~Zhan Lum, Zhou Ye, Xiaoguang Dong, Hamid Marvi, Onder Erin, Wenqi Hu, and Metin Sitti.
\newblock Shape-programmable magnetic soft matter.
\newblock \emph{Proceedings of the National Academy of Sciences}, 113\penalty0 (41), October 2016.
\newblock ISSN 0027-8424, 1091-6490.
\newblock \doi{10.1073/pnas.1608193113}.
\newblock URL \url{https://pnas.org/doi/full/10.1073/pnas.1608193113}.

\bibitem[Ren et~al.(2019)Ren, Hu, Dong, and Sitti]{ren_multi-functional_2019}
Ziyu Ren, Wenqi Hu, Xiaoguang Dong, and Metin Sitti.
\newblock Multi-functional soft-bodied jellyfish-like swimming.
\newblock \emph{Nature Communications}, 10\penalty0 (1):\penalty0 2703, July 2019.
\newblock ISSN 2041-1723.
\newblock \doi{10.1038/s41467-019-10549-7}.
\newblock URL \url{https://www.nature.com/articles/s41467-019-10549-7}.

\bibitem[Ebbens and Howse(2010)]{ebbens_pursuit_2010}
Stephen~J. Ebbens and Jonathan~R. Howse.
\newblock In pursuit of propulsion at the nanoscale.
\newblock \emph{Soft Matter}, 6\penalty0 (4):\penalty0 726, 2010.
\newblock ISSN 1744-683X, 1744-6848.
\newblock \doi{10.1039/b918598d}.
\newblock URL \url{https://xlink.rsc.org/?DOI=b918598d}.

\bibitem[Venugopalan et~al.(2020)Venugopalan, Esteban-Fernández De~Ávila, Pal, Ghosh, and Wang]{venugopalan_fantastic_2020}
Pooyath~Lekshmy Venugopalan, Berta Esteban-Fernández De~Ávila, Malay Pal, Ambarish Ghosh, and Joseph Wang.
\newblock Fantastic {Voyage} of {Nanomotors} into the {Cell}.
\newblock \emph{ACS Nano}, 14\penalty0 (8):\penalty0 9423--9439, August 2020.
\newblock ISSN 1936-0851, 1936-086X.
\newblock \doi{10.1021/acsnano.0c05217}.
\newblock URL \url{https://pubs.acs.org/doi/10.1021/acsnano.0c05217}.

\bibitem[Gao et~al.(2012)Gao, Kagan, Pak, Clawson, Campuzano, Chuluun‐Erdene, Shipton, Fullerton, Zhang, Lauga, and Wang]{gao_cargotowing_2012}
Wei Gao, Daniel Kagan, On~Shun Pak, Corbin Clawson, Susana Campuzano, Erdembileg Chuluun‐Erdene, Erik Shipton, Eric~E. Fullerton, Liangfang Zhang, Eric Lauga, and Joseph Wang.
\newblock Cargo‐{Towing} {Fuel}‐{Free} {Magnetic} {Nanoswimmers} for {Targeted} {Drug} {Delivery}.
\newblock \emph{Small}, 8\penalty0 (3):\penalty0 460--467, February 2012.
\newblock ISSN 1613-6810, 1613-6829.
\newblock \doi{10.1002/smll.201101909}.
\newblock URL \url{https://onlinelibrary.wiley.com/doi/10.1002/smll.201101909}.

\bibitem[Sundararajan et~al.(2008)Sundararajan, Lammert, Zudans, Crespi, and Sen]{sundararajan_catalytic_2008}
Shakuntala Sundararajan, Paul~E. Lammert, Andrew~W. Zudans, Vincent~H. Crespi, and Ayusman Sen.
\newblock Catalytic {Motors} for {Transport} of {Colloidal} {Cargo}.
\newblock \emph{Nano Letters}, 8\penalty0 (5):\penalty0 1271--1276, May 2008.
\newblock ISSN 1530-6984, 1530-6992.
\newblock \doi{10.1021/nl072275j}.
\newblock URL \url{https://pubs.acs.org/doi/10.1021/nl072275j}.

\bibitem[Paxton et~al.(2006)Paxton, Sundararajan, Mallouk, and Sen]{paxton_chemical_2006}
Walter~F. Paxton, Shakuntala Sundararajan, Thomas~E. Mallouk, and Ayusman Sen.
\newblock Chemical {Locomotion}.
\newblock \emph{Angewandte Chemie International Edition}, 45\penalty0 (33):\penalty0 5420--5429, August 2006.
\newblock ISSN 1433-7851, 1521-3773.
\newblock \doi{10.1002/anie.200600060}.
\newblock URL \url{https://onlinelibrary.wiley.com/doi/10.1002/anie.200600060}.

\bibitem[Sanchez et~al.(2011)Sanchez, Welch, Nicastro, and Dogic]{sanchez_cilia-like_2011}
Timothy Sanchez, David Welch, Daniela Nicastro, and Zvonimir Dogic.
\newblock Cilia-{Like} {Beating} of {Active} {Microtubule} {Bundles}.
\newblock \emph{Science}, 333\penalty0 (6041):\penalty0 456--459, July 2011.
\newblock ISSN 0036-8075, 1095-9203.
\newblock \doi{10.1126/science.1203963}.
\newblock URL \url{https://www.science.org/doi/10.1126/science.1203963}.

\bibitem[Gao and Wang(2014)]{gao_synthetic_2014}
Wei Gao and Joseph Wang.
\newblock Synthetic micro/nanomotors in drug delivery.
\newblock \emph{Nanoscale}, 6\penalty0 (18):\penalty0 10486--10494, 2014.
\newblock ISSN 2040-3364, 2040-3372.
\newblock \doi{10.1039/C4NR03124E}.
\newblock URL \url{https://xlink.rsc.org/?DOI=C4NR03124E}.

\bibitem[Hawkes et~al.(2010)Hawkes, An, Benbernou, Tanaka, Kim, Demaine, Rus, and Wood]{hawkes_programmable_2010}
E.~Hawkes, B.~An, N.~M. Benbernou, H.~Tanaka, S.~Kim, E.~D. Demaine, D.~Rus, and R.~J. Wood.
\newblock Programmable matter by folding.
\newblock \emph{Proceedings of the National Academy of Sciences}, 107\penalty0 (28):\penalty0 12441--12445, July 2010.
\newblock ISSN 0027-8424, 1091-6490.
\newblock \doi{10.1073/pnas.0914069107}.
\newblock URL \url{https://pnas.org/doi/full/10.1073/pnas.0914069107}.

\bibitem[Zhou et~al.(2021)Zhou, Mayorga-Martinez, Pané, Zhang, and Pumera]{zhou_magnetically_2021}
Huaijuan Zhou, Carmen~C. Mayorga-Martinez, Salvador Pané, Li~Zhang, and Martin Pumera.
\newblock Magnetically {Driven} {Micro} and {Nanorobots}.
\newblock \emph{Chemical Reviews}, 121\penalty0 (8):\penalty0 4999--5041, April 2021.
\newblock ISSN 0009-2665, 1520-6890.
\newblock \doi{10.1021/acs.chemrev.0c01234}.
\newblock URL \url{https://pubs.acs.org/doi/10.1021/acs.chemrev.0c01234}.

\bibitem[Servant et~al.(2015)Servant, Qiu, Mazza, Kostarelos, and Nelson]{servant_controlled_2015}
Ania Servant, Famin Qiu, Mariarosa Mazza, Kostas Kostarelos, and Bradley~J. Nelson.
\newblock Controlled {In} {Vivo} {Swimming} of a {Swarm} of {Bacteria}‐{Like} {Microrobotic} {Flagella}.
\newblock \emph{Advanced Materials}, 27\penalty0 (19):\penalty0 2981--2988, May 2015.
\newblock ISSN 0935-9648, 1521-4095.
\newblock \doi{10.1002/adma.201404444}.
\newblock URL \url{https://onlinelibrary.wiley.com/doi/10.1002/adma.201404444}.

\bibitem[Chen et~al.(2018)Chen, Chang, Teymourian, Lu, Li, He, Fang, Liang, Mou, Guan, and Wang]{chen_bioinspired_2018}
Chuanrui Chen, Xiaocong Chang, Hazhir Teymourian, Xiaolong Lu, Jinxing Li, Sha He, Chengcheng Fang, Yuyan Liang, Fangzhi Mou, Jianguo Guan, and Joseph Wang.
\newblock Bioinspired {Chemical} {Communication} between {Synthetic} {Nanomotors}.
\newblock \emph{Angewandte Chemie International Edition}, 57\penalty0 (1):\penalty0 241--245, January 2018.
\newblock ISSN 1433-7851, 1521-3773.
\newblock \doi{10.1002/anie.201710376}.
\newblock URL \url{https://onlinelibrary.wiley.com/doi/10.1002/anie.201710376}.

\bibitem[Miskin et~al.(2020)Miskin, Cortese, Dorsey, Esposito, Reynolds, Liu, Cao, Muller, McEuen, and Cohen]{miskin_electronically_2020}
Marc~Z. Miskin, Alejandro~J. Cortese, Kyle Dorsey, Edward~P. Esposito, Michael~F. Reynolds, Qingkun Liu, Michael Cao, David~A. Muller, Paul~L. McEuen, and Itai Cohen.
\newblock Electronically integrated, mass-manufactured, microscopic robots.
\newblock \emph{Nature}, 584\penalty0 (7822):\penalty0 557--561, August 2020.
\newblock ISSN 0028-0836, 1476-4687.
\newblock \doi{10.1038/s41586-020-2626-9}.
\newblock URL \url{https://www.nature.com/articles/s41586-020-2626-9}.

\bibitem[Falk et~al.(2021)Falk, Alizadehyazdi, Jaeger, and Murugan]{falk_learning_2021}
Martin~J. Falk, Vahid Alizadehyazdi, Heinrich Jaeger, and Arvind Murugan.
\newblock Learning to control active matter.
\newblock \emph{Physical Review Research}, 3\penalty0 (3):\penalty0 033291, September 2021.
\newblock ISSN 2643-1564.
\newblock \doi{10.1103/PhysRevResearch.3.033291}.
\newblock URL \url{https://link.aps.org/doi/10.1103/PhysRevResearch.3.033291}.

\bibitem[Cichos et~al.(2020)Cichos, Gustavsson, Mehlig, and Volpe]{cichos_machine_2020}
Frank Cichos, Kristian Gustavsson, Bernhard Mehlig, and Giovanni Volpe.
\newblock Machine learning for active matter.
\newblock \emph{Nature Machine Intelligence}, 2\penalty0 (2):\penalty0 94--103, February 2020.
\newblock ISSN 2522-5839.
\newblock \doi{10.1038/s42256-020-0146-9}.
\newblock URL \url{https://www.nature.com/articles/s42256-020-0146-9}.

\bibitem[Pollard and Cooper(2009)]{pollard_actin_2009}
Thomas~D. Pollard and John~A. Cooper.
\newblock Actin, a {Central} {Player} in {Cell} {Shape} and {Movement}.
\newblock \emph{Science}, 326\penalty0 (5957):\penalty0 1208--1212, November 2009.
\newblock ISSN 0036-8075, 1095-9203.
\newblock \doi{10.1126/science.1175862}.
\newblock URL \url{https://www.science.org/doi/10.1126/science.1175862}.

\bibitem[Fletcher and Mullins(2010)]{fletcher_cell_2010}
Daniel~A. Fletcher and R.~Dyche Mullins.
\newblock Cell mechanics and the cytoskeleton.
\newblock \emph{Nature}, 463\penalty0 (7280):\penalty0 485--492, January 2010.
\newblock ISSN 0028-0836, 1476-4687.
\newblock \doi{10.1038/nature08908}.
\newblock URL \url{https://www.nature.com/articles/nature08908}.

\bibitem[Cheng et~al.(2020)Cheng, Felix, and Othmer]{cheng_roles_2020}
Yougan Cheng, Bryan Felix, and Hans~G. Othmer.
\newblock The {Roles} of {Signaling} in {Cytoskeletal} {Changes}, {Random} {Movement}, {Direction}-{Sensing} and {Polarization} of {Eukaryotic} {Cells}.
\newblock \emph{Cells}, 9\penalty0 (6):\penalty0 1437, June 2020.
\newblock ISSN 2073-4409.
\newblock \doi{10.3390/cells9061437}.
\newblock URL \url{https://www.mdpi.com/2073-4409/9/6/1437}.

\bibitem[Boudet et~al.(2021)Boudet, Lintuvuori, Lacouture, Barois, Deblais, Xie, Cassagnere, Tregon, Brückner, Baret, and Kellay]{boudet_collections_2021}
J.~F. Boudet, J.~Lintuvuori, C.~Lacouture, T.~Barois, A.~Deblais, K.~Xie, S.~Cassagnere, B.~Tregon, D.~B. Brückner, J.~C. Baret, and H.~Kellay.
\newblock From collections of independent, mindless robots to flexible, mobile, and directional superstructures.
\newblock \emph{Science Robotics}, 6\penalty0 (56):\penalty0 eabd0272, July 2021.
\newblock ISSN 2470-9476.
\newblock \doi{10.1126/scirobotics.abd0272}.
\newblock URL \url{https://www.science.org/doi/10.1126/scirobotics.abd0272}.

\bibitem[Savoie et~al.(2019)Savoie, Berrueta, Jackson, Pervan, Warkentin, Li, Murphey, Wiesenfeld, and Goldman]{savoie_robot_2019}
William Savoie, Thomas~A. Berrueta, Zachary Jackson, Ana Pervan, Ross Warkentin, Shengkai Li, Todd~D. Murphey, Kurt Wiesenfeld, and Daniel~I. Goldman.
\newblock A robot made of robots: {Emergent} transport and control of a smarticle ensemble.
\newblock \emph{Science Robotics}, 4\penalty0 (34):\penalty0 eaax4316, September 2019.
\newblock ISSN 2470-9476.
\newblock \doi{10.1126/scirobotics.aax4316}.
\newblock URL \url{https://www.science.org/doi/10.1126/scirobotics.aax4316}.

\bibitem[Veenstra et~al.(2025)Veenstra, Scheibner, Brandenbourger, Binysh, Souslov, Vitelli, and Coulais]{veenstra_adaptive_2025}
Jonas Veenstra, Colin Scheibner, Martin Brandenbourger, Jack Binysh, Anton Souslov, Vincenzo Vitelli, and Corentin Coulais.
\newblock Adaptive locomotion of active solids.
\newblock \emph{Nature}, March 2025.
\newblock ISSN 0028-0836, 1476-4687.
\newblock \doi{10.1038/s41586-025-08646-3}.
\newblock URL \url{https://www.nature.com/articles/s41586-025-08646-3}.

\bibitem[Vutukuri et~al.(2020)Vutukuri, Hoore, Abaurrea-Velasco, Van~Buren, Dutto, Auth, Fedosov, Gompper, and Vermant]{vutukuri_active_2020}
Hanumantha~Rao Vutukuri, Masoud Hoore, Clara Abaurrea-Velasco, Lennard Van~Buren, Alessandro Dutto, Thorsten Auth, Dmitry~A. Fedosov, Gerhard Gompper, and Jan Vermant.
\newblock Active particles induce large shape deformations in giant lipid vesicles.
\newblock \emph{Nature}, 586\penalty0 (7827):\penalty0 52--56, October 2020.
\newblock ISSN 0028-0836, 1476-4687.
\newblock \doi{10.1038/s41586-020-2730-x}.
\newblock URL \url{https://www.nature.com/articles/s41586-020-2730-x}.

\bibitem[Le~Nagard et~al.(2022)Le~Nagard, Brown, Dawson, Martinez, Poon, and Staykova]{le_nagard_encapsulated_2022}
Lucas Le~Nagard, Aidan~T. Brown, Angela Dawson, Vincent~A. Martinez, Wilson C.~K. Poon, and Margarita Staykova.
\newblock Encapsulated bacteria deform lipid vesicles into flagellated swimmers.
\newblock \emph{Proceedings of the National Academy of Sciences}, 119\penalty0 (34):\penalty0 e2206096119, August 2022.
\newblock ISSN 0027-8424, 1091-6490.
\newblock \doi{10.1073/pnas.2206096119}.
\newblock URL \url{https://pnas.org/doi/full/10.1073/pnas.2206096119}.

\bibitem[Ramos et~al.(2020)Ramos, Cordero, and Soto]{ramos_bacteria_2020}
Gabriel Ramos, María~Luisa Cordero, and Rodrigo Soto.
\newblock Bacteria driving droplets.
\newblock \emph{Soft Matter}, 16\penalty0 (5):\penalty0 1359--1365, 2020.
\newblock ISSN 1744-683X, 1744-6848.
\newblock \doi{10.1039/C9SM01839E}.
\newblock URL \url{http://xlink.rsc.org/?DOI=C9SM01839E}.

\bibitem[Heinrich et~al.(1999)Heinrich, Bo{\v{z}}i{\v{c}}, Svetina, and {\v{Z}}ek{\v{s}}]{heinrich1999vesicle}
Volkmar Heinrich, Bojan Bo{\v{z}}i{\v{c}}, Sa{\v{s}}a Svetina, and Bo{\v{s}}tjan {\v{Z}}ek{\v{s}}.
\newblock Vesicle deformation by an axial load: from elongated shapes to tethered vesicles.
\newblock \emph{Biophysical journal}, 76\penalty0 (4):\penalty0 2056--2071, 1999.

\bibitem[Schönhöfer and Glotzer(2025)]{schonhofer_collective_2025}
Philipp W.~A. Schönhöfer and Sharon~C. Glotzer.
\newblock Collective behavior of “flexicles”.
\newblock \emph{Proceedings of the National Academy of Sciences}, 122\penalty0 (36):\penalty0 e2426850122, 2025.
\newblock \doi{10.1073/pnas.2426850122}.
\newblock URL \url{https://www.pnas.org/doi/abs/10.1073/pnas.2426850122}.

\bibitem[Lee et~al.(2023)Lee, Schönhöfer, and Glotzer]{lee_complex_2023}
Sophie~Y. Lee, Philipp W.~A. Schönhöfer, and Sharon~C. Glotzer.
\newblock Complex motion of steerable vesicular robots filled with active colloidal rods.
\newblock \emph{Scientific Reports}, 13\penalty0 (1):\penalty0 22773, December 2023.
\newblock ISSN 2045-2322.
\newblock \doi{10.1038/s41598-023-49314-8}.
\newblock URL \url{https://www.nature.com/articles/s41598-023-49314-8}.

\bibitem[Paoluzzi et~al.(2016)Paoluzzi, Di~Leonardo, Marchetti, and Angelani]{paoluzzi_shape_2016}
Matteo Paoluzzi, Roberto Di~Leonardo, M.~Cristina Marchetti, and Luca Angelani.
\newblock Shape and {Displacement} {Fluctuations} in {Soft} {Vesicles} {Filled} by {Active} {Particles}.
\newblock \emph{Scientific Reports}, 6\penalty0 (1):\penalty0 34146, December 2016.
\newblock ISSN 2045-2322.
\newblock \doi{10.1038/srep34146}.
\newblock URL \url{http://www.nature.com/articles/srep34146}.

\bibitem[Schlick(2010)]{schlick_molecular_2010}
Tamar Schlick.
\newblock \emph{Molecular {Modeling} and {Simulation}: {An} {Interdisciplinary} {Guide}: {An} {Interdisciplinary} {Guide}}, volume~21 of \emph{Interdisciplinary {Applied} {Mathematics}}.
\newblock Springer New York, New York, NY, 2010.
\newblock ISBN 978-1-4419-6350-5 978-1-4419-6351-2.
\newblock \doi{10.1007/978-1-4419-6351-2}.
\newblock URL \url{https://link.springer.com/10.1007/978-1-4419-6351-2}.

\bibitem[Von~Smoluchowski(1906)]{von_smoluchowski_zur_1906}
M.~Von~Smoluchowski.
\newblock Zur kinetischen {Theorie} der {Brownschen} {Molekularbewegung} und der {Suspensionen}.
\newblock \emph{Annalen der Physik}, 326\penalty0 (14):\penalty0 756--780, January 1906.
\newblock ISSN 0003-3804, 1521-3889.
\newblock \doi{10.1002/andp.19063261405}.
\newblock URL \url{https://onlinelibrary.wiley.com/doi/10.1002/andp.19063261405}.

\bibitem[Weeks et~al.(1971)Weeks, Chandler, and Andersen]{weeks_role_1971}
John~D. Weeks, David Chandler, and Hans~C. Andersen.
\newblock Role of {Repulsive} {Forces} in {Determining} the {Equilibrium} {Structure} of {Simple} {Liquids}.
\newblock \emph{The Journal of Chemical Physics}, 54\penalty0 (12):\penalty0 5237--5247, June 1971.
\newblock ISSN 0021-9606, 1089-7690.
\newblock \doi{10.1063/1.1674820}.
\newblock URL \url{https://pubs.aip.org/jcp/article/54/12/5237/85492/Role-of-Repulsive-Forces-in-Determining-the}.

\bibitem[Noguchi and Gompper(2005)]{noguchi_dynamics_2005}
Hiroshi Noguchi and Gerhard Gompper.
\newblock Dynamics of fluid vesicles in shear flow: {Effect} of membrane viscosity and thermal fluctuations.
\newblock \emph{Physical Review E}, 72\penalty0 (1):\penalty0 011901, July 2005.
\newblock ISSN 1539-3755, 1550-2376.
\newblock \doi{10.1103/PhysRevE.72.011901}.
\newblock URL \url{https://link.aps.org/doi/10.1103/PhysRevE.72.011901}.

\bibitem[Helfrich(1973)]{helfrich_elastic_1973}
W~Helfrich.
\newblock Elastic {Properties} of {Lipid} {Bilayers}: {Theory} and {Possible} {Experiments}.
\newblock \emph{Zeitschrift für Naturforschung C}, 28\penalty0 (11-12):\penalty0 693--703, December 1973.
\newblock ISSN 1865-7125, 0939-5075.
\newblock \doi{10.1515/znc-1973-11-1209}.
\newblock URL \url{https://www.degruyter.com/document/doi/10.1515/znc-1973-11-1209/html}.

\bibitem[Gompper and Kroll(1996)]{gompper_random_1996}
G.~Gompper and D.~M. Kroll.
\newblock Random {Surface} {Discretizations} and the {Renormalization} of the {Bending} {Rigidity}.
\newblock \emph{Journal de Physique I}, 6\penalty0 (10):\penalty0 1305--1320, October 1996.
\newblock ISSN 1155-4304, 1286-4862.
\newblock \doi{10.1051/jp1:1996246}.
\newblock URL \url{http://www.edpsciences.org/10.1051/jp1:1996246}.

\bibitem[Nelson et~al.(2004)Nelson, Piran, and Weinberg]{nelson_statistical_2004}
David~R. Nelson, Tsvi Piran, and Steven Weinberg.
\newblock \emph{Statistical mechanics of membranes and surfaces}.
\newblock World Scientific, Singapore, 2nd ed edition, 2004.
\newblock ISBN 978-981-238-760-8.

\bibitem[Gompper and Kroll(1997)]{gompper_network_1997}
G~Gompper and D~M Kroll.
\newblock Network models of fluid, hexatic and polymerized membranes.
\newblock \emph{Journal of Physics: Condensed Matter}, 9\penalty0 (42):\penalty0 8795--8834, October 1997.
\newblock ISSN 0953-8984, 1361-648X.
\newblock \doi{10.1088/0953-8984/9/42/001}.
\newblock URL \url{https://iopscience.iop.org/article/10.1088/0953-8984/9/42/001}.

\bibitem[Anderson et~al.(2020)Anderson, Glaser, and Glotzer]{anderson_hoomd-blue_2020}
Joshua~A. Anderson, Jens Glaser, and Sharon~C. Glotzer.
\newblock {HOOMD}-blue: {A} {Python} package for high-performance molecular dynamics and hard particle {Monte} {Carlo} simulations.
\newblock \emph{Computational Materials Science}, 173:\penalty0 109363, February 2020.
\newblock ISSN 09270256.
\newblock \doi{10.1016/j.commatsci.2019.109363}.
\newblock URL \url{https://linkinghub.elsevier.com/retrieve/pii/S0927025619306627}.

\bibitem[Ramasubramani et~al.(2020)Ramasubramani, Dice, Harper, Spellings, Anderson, and Glotzer]{ramasubramani_freud_2020}
Vyas Ramasubramani, Bradley~D. Dice, Eric~S. Harper, Matthew~P. Spellings, Joshua~A. Anderson, and Sharon~C. Glotzer.
\newblock freud: {A} software suite for high throughput analysis of particle simulation data.
\newblock \emph{Computer Physics Communications}, 254:\penalty0 107275, September 2020.
\newblock ISSN 00104655.
\newblock \doi{10.1016/j.cpc.2020.107275}.
\newblock URL \url{https://linkinghub.elsevier.com/retrieve/pii/S0010465520300916}.

\bibitem[Adorf et~al.(2018)Adorf, Dodd, Ramasubramani, and Glotzer]{adorf_simple_2018}
Carl~S. Adorf, Paul~M. Dodd, Vyas Ramasubramani, and Sharon~C. Glotzer.
\newblock Simple data and workflow management with the signac framework.
\newblock \emph{Computational Materials Science}, 146:\penalty0 220--229, April 2018.
\newblock ISSN 09270256.
\newblock \doi{10.1016/j.commatsci.2018.01.035}.
\newblock URL \url{https://linkinghub.elsevier.com/retrieve/pii/S0927025618300429}.

\bibitem[Abaurrea-Velasco et~al.(2019)Abaurrea-Velasco, Auth, and Gompper]{abaurrea-velasco_vesicles_2019}
Clara Abaurrea-Velasco, Thorsten Auth, and Gerhard Gompper.
\newblock Vesicles with internal active filaments: self-organized propulsion controls shape, motility, and dynamical response.
\newblock \emph{New Journal of Physics}, 21\penalty0 (12):\penalty0 123024, December 2019.
\newblock ISSN 1367-2630.
\newblock \doi{10.1088/1367-2630/ab5c70}.
\newblock URL \url{https://iopscience.iop.org/article/10.1088/1367-2630/ab5c70}.

\bibitem[Greene et~al.(2016)Greene, Henderson, Gomez, Paxton, VanDelinder, and Bachand]{greene_role_2016}
Adrienne~C. Greene, Ian~M. Henderson, Andrew Gomez, Walter~F. Paxton, Virginia VanDelinder, and George~D. Bachand.
\newblock The {Role} of {Membrane} {Fluidization} in the {Gel}-{Assisted} {Formation} of {Giant} {Polymersomes}.
\newblock \emph{PLOS ONE}, 11\penalty0 (7):\penalty0 e0158729, July 2016.
\newblock ISSN 1932-6203.
\newblock \doi{10.1371/journal.pone.0158729}.
\newblock URL \url{https://dx.plos.org/10.1371/journal.pone.0158729}.

\bibitem[Sarangapani and Asbury(2014)]{sarangapani_catch_2014}
Krishna~K. Sarangapani and Charles~L. Asbury.
\newblock Catch and release: how do kinetochores hook the right microtubules during mitosis?
\newblock \emph{Trends in Genetics}, 30\penalty0 (4):\penalty0 150--159, April 2014.
\newblock ISSN 01689525.
\newblock \doi{10.1016/j.tig.2014.02.004}.
\newblock URL \url{https://linkinghub.elsevier.com/retrieve/pii/S0168952514000274}.

\bibitem[Berleman and Kirby(2009)]{berleman_deciphering_2009}
James~E. Berleman and John~R. Kirby.
\newblock Deciphering the hunting strategy of a bacterial wolfpack.
\newblock \emph{FEMS Microbiology Reviews}, 33\penalty0 (5):\penalty0 942--957, September 2009.
\newblock ISSN 1574-6976.
\newblock \doi{10.1111/j.1574-6976.2009.00185.x}.
\newblock URL \url{https://academic.oup.com/femsre/article-lookup/doi/10.1111/j.1574-6976.2009.00185.x}.

\bibitem[Gordon(2010)]{gordon_interaction_2010}
Deborah~M. Gordon.
\newblock \emph{Interaction {Networks} and {Colony} {Behavior}}.
\newblock Princeton University Press, Princeton, 2010.
\newblock ISBN 978-1-4008-3544-7.
\newblock \doi{doi:10.1515/9781400835447}.
\newblock URL \url{https://doi.org/10.1515/9781400835447}.

\bibitem[Tarazona et~al.(2013)Tarazona, Chacón, and Bresme]{tarazona_thermal_2013}
Pedro Tarazona, Enrique Chacón, and Fernando Bresme.
\newblock Thermal fluctuations and bending rigidity of bilayer membranes.
\newblock \emph{The Journal of Chemical Physics}, 139\penalty0 (9):\penalty0 094902, September 2013.
\newblock ISSN 0021-9606, 1089-7690.
\newblock \doi{10.1063/1.4818421}.
\newblock URL \url{https://pubs.aip.org/jcp/article/139/9/094902/315645/Thermal-fluctuations-and-bending-rigidity-of}.

\bibitem[Rawicz et~al.(2000)Rawicz, Olbrich, McIntosh, Needham, and Evans]{rawicz_effect_2000}
W.~Rawicz, K.C. Olbrich, T.~McIntosh, D.~Needham, and E.~Evans.
\newblock Effect of {Chain} {Length} and {Unsaturation} on {Elasticity} of {Lipid} {Bilayers}.
\newblock \emph{Biophysical Journal}, 79\penalty0 (1):\penalty0 328--339, July 2000.
\newblock ISSN 0006-3495.
\newblock \doi{10.1016/s0006-3495(00)76295-3}.
\newblock URL \url{https://linkinghub.elsevier.com/retrieve/pii/S0006349500762953}.
\newblock Publisher: Elsevier BV.

\bibitem[Bermúdez et~al.(2004)Bermúdez, Hammer, and Discher]{bermudez_effect_2004}
H.~Bermúdez, D.~A. Hammer, and D.~E. Discher.
\newblock Effect of {Bilayer} {Thickness} on {Membrane} {Bending} {Rigidity}.
\newblock \emph{Langmuir}, 20\penalty0 (3):\penalty0 540--543, February 2004.
\newblock ISSN 0743-7463, 1520-5827.
\newblock \doi{10.1021/la035497f}.
\newblock URL \url{https://pubs.acs.org/doi/10.1021/la035497f}.
\newblock Publisher: American Chemical Society (ACS).

\bibitem[Doole et~al.(2022)Doole, Kumarage, Ashkar, and Brown]{doole_cholesterol_2022}
Fathima~T. Doole, Teshani Kumarage, Rana Ashkar, and Michael~F. Brown.
\newblock Cholesterol {Stiffening} of {Lipid} {Membranes}.
\newblock \emph{The Journal of Membrane Biology}, 255\penalty0 (4-5):\penalty0 385--405, October 2022.
\newblock ISSN 0022-2631, 1432-1424.
\newblock \doi{10.1007/s00232-022-00263-9}.
\newblock URL \url{https://link.springer.com/10.1007/s00232-022-00263-9}.

\bibitem[Faizi et~al.(2019)Faizi, Frey, Steinkühler, Dimova, and Vlahovska]{faizi_bending_2019}
Hammad~A. Faizi, Shelli~L. Frey, Jan Steinkühler, Rumiana Dimova, and Petia~M. Vlahovska.
\newblock Bending rigidity of charged lipid bilayer membranes.
\newblock \emph{Soft Matter}, 15\penalty0 (29):\penalty0 6006--6013, 2019.
\newblock ISSN 1744-683X, 1744-6848.
\newblock \doi{10.1039/C9SM00772E}.
\newblock URL \url{https://xlink.rsc.org/?DOI=C9SM00772E}.

\bibitem[Fowler et~al.(2016)Fowler, Hélie, Duncan, Chavent, Koldsø, and Sansom]{fowler_membrane_2016}
Philip~W. Fowler, Jean Hélie, Anna Duncan, Matthieu Chavent, Heidi Koldsø, and Mark S.~P. Sansom.
\newblock Membrane stiffness is modified by integral membrane proteins.
\newblock \emph{Soft Matter}, 12\penalty0 (37):\penalty0 7792--7803, 2016.
\newblock ISSN 1744-683X, 1744-6848.
\newblock \doi{10.1039/C6SM01186A}.
\newblock URL \url{https://xlink.rsc.org/?DOI=C6SM01186A}.

\bibitem[LoPresti et~al.(2009)LoPresti, Lomas, Massignani, Smart, and Battaglia]{lopresti_polymersomes_2009}
Caterina LoPresti, Hannah Lomas, Marzia Massignani, Thomas Smart, and Giuseppe Battaglia.
\newblock Polymersomes: nature inspired nanometer sized compartments.
\newblock \emph{Journal of Materials Chemistry}, 19\penalty0 (22):\penalty0 3576, 2009.
\newblock ISSN 0959-9428, 1364-5501.
\newblock \doi{10.1039/b818869f}.
\newblock URL \url{https://xlink.rsc.org/?DOI=b818869f}.
\newblock Publisher: Royal Society of Chemistry (RSC).

\bibitem[Lasic(1994)]{lasic_sterically_1994}
Danilo~D. Lasic.
\newblock Sterically {Stabilized} {Vesicles}.
\newblock \emph{Angewandte Chemie International Edition in English}, 33\penalty0 (17):\penalty0 1685--1698, September 1994.
\newblock ISSN 0570-0833.
\newblock \doi{10.1002/anie.199416851}.
\newblock URL \url{https://onlinelibrary.wiley.com/doi/10.1002/anie.199416851}.
\newblock Publisher: Wiley.

\bibitem[Faizi et~al.(2022)Faizi, Tsui, Dimova, and Vlahovska]{faizi_bending_2022}
Hammad~A. Faizi, Annie Tsui, Rumiana Dimova, and Petia~M. Vlahovska.
\newblock Bending {Rigidity}, {Capacitance}, and {Shear} {Viscosity} of {Giant} {Vesicle} {Membranes} {Prepared} by {Spontaneous} {Swelling}, {Electroformation}, {Gel}-{Assisted}, and {Phase} {Transfer} {Methods}: {A} {Comparative} {Study}.
\newblock \emph{Langmuir}, 38\penalty0 (34):\penalty0 10548--10557, August 2022.
\newblock ISSN 0743-7463, 1520-5827.
\newblock \doi{10.1021/acs.langmuir.2c01402}.
\newblock URL \url{https://pubs.acs.org/doi/10.1021/acs.langmuir.2c01402}.
\newblock Publisher: American Chemical Society (ACS).

\bibitem[Rideau et~al.(2018)Rideau, Dimova, Schwille, Wurm, and Landfester]{rideau_liposomes_2018}
Emeline Rideau, Rumiana Dimova, Petra Schwille, Frederik~R. Wurm, and Katharina Landfester.
\newblock Liposomes and polymersomes: a comparative review towards cell mimicking.
\newblock \emph{Chemical Society Reviews}, 47\penalty0 (23):\penalty0 8572--8610, 2018.
\newblock ISSN 0306-0012, 1460-4744.
\newblock \doi{10.1039/c8cs00162f}.
\newblock URL \url{https://xlink.rsc.org/?DOI=C8CS00162F}.
\newblock Publisher: Royal Society of Chemistry (RSC).

\bibitem[Discher et~al.(1999)Discher, Won, Ege, Lee, Bates, Discher, and Hammer]{discher_polymersomes_1999}
Bohdana~M. Discher, You-Yeon Won, David~S. Ege, James C-M. Lee, Frank~S. Bates, Dennis~E. Discher, and Daniel~A. Hammer.
\newblock Polymersomes: {Tough} {Vesicles} {Made} from {Diblock} {Copolymers}.
\newblock \emph{Science}, 284\penalty0 (5417):\penalty0 1143--1146, May 1999.
\newblock ISSN 0036-8075, 1095-9203.
\newblock \doi{10.1126/science.284.5417.1143}.
\newblock URL \url{https://www.science.org/doi/10.1126/science.284.5417.1143}.
\newblock Publisher: American Association for the Advancement of Science (AAAS).

\end{thebibliography}
\bibliographystyle{unsrtnat}
\setcitestyle{numbers,square}

\pagebreak
\widetext
\widetext
\begin{center}
  \textbf{\large Supplemental Materials:\\ Emergent Microrobotic Behavior of Active Flexicles in Complex Environments}
\end{center}

\renewcommand{\thefigure}{S\arabic{figure}}
\setcounter{figure}{0} 


\begin{table}[b]
\caption{\label{tab:units}%
Estimated physical units corresponding to the model parameters used in our simulations. The conversion is based on mapping the dimensionless simulation units for mass, length, energy, and time to representative physical values, allowing for comparison with experimentally relevant systems.}
\begin{ruledtabular}
\begin{tabular}{lcc}
\textrm{Parameters}&
\textrm{Model units}&
\textrm{Physical units}\\
\colrule
mass scale $m$ & 1 & $7 \times 10^{-15}\,\mathrm{kg}$\\
Length scale $\sigma$ & 1 & $1\,\mu\mathrm{m}$ \\
Time scale $\tau$ & $1$ & $5.8 \times 10^{-4}\,\mathrm{s}$ \\
Thermal energy $k_BT$ & 0.2 & $4.14 \times 10^{-21}\,\mathrm{J}$ \\
SP rod length & 3 & $3\,\mu\mathrm{m}$ \\
SP rod P\'eclet number $\mathrm{Pe}$ & 100 & 100 \\
drag coefficient $\gamma$ & 250 $m/\tau$ & $1.85 \times 10^{-8}\,\mathrm{kg}/\mathrm{s}$ \\
Vesicle equilibrium radius $R$ & 8 & $8\,\mu\mathrm{m}$ \\
Bending rigidity $\kappa_{H}$ & {$50$--$1000$} $k_BT$ & {$2$--$200$} $\times (10^{-19})\,\mathrm{J}$\\
Bond stiffness $\kappa_{B}$ & $1330k_BT$ & $5.508$ $\times (10^{-18})\,\mathrm{J}$\\

\end{tabular}
\end{ruledtabular}
\end{table}
\pagebreak

\subsection{Electronic Supplementary Information}
\label{sec:vid}
\begin{enumerate}[align=left, leftmargin=*]
  \item[Movie 1:] A flexicle with $\kappa_H=100\,k_BT$ encounters a fixed spherical object with radius $R_{\text{sph}}=1.3\,R_\text{flex}$ and latches on the surface 
  \item[Movie 2:] A flexicle with $\kappa_H=10000\,k_BT$ latches onto a movable, passive spherical object with radius $R_{\text{sph}}=2.75\,R_\text{flex}$ and transports it linearly.
  \item[Movie 3:] A flexicle with $\kappa_H=100\,k_BT$ encounters a fixed cylindrical object with radius $R_\text{cyl}=0.75\,R_\text{flex}$ and latches on the surface 
  \item[Movie 4:] A flexicle with $\kappa_H=100\,k_BT$ adhered to the floor of a square box with walls of height $h_\text{wall}=1.75\,R_\text{flex}$ crawls across the floor, reaches the wall edge, navigates toward a corner, and eventually escapes after several failed attempts. 
  \item[Movie 5:] A flexicle with $\kappa_H=100\,k_BT$ subjected to gravity climbs a staircase with step height $h_{\text{step}}=R_\text{flex}$. 
  \item[Movie 6:] A flexicle with $\kappa_H = 100\,k_BT$ latches onto a movable spherical object with radius $R_\text{sph}=2.75\,R_\text{flex}$ and initially oscillates along the surface. As the membrane bending rigidity is gradually increased to $\kappa_H = 10000\,k_BT$, the flexicle starts to transport the object linearly. 
  \item[Movie 7:] A flexicle with $\kappa_H = 10000\,k_BT$ latches onto a movable spherical object with radius $R_{\text{sph}}=2.75\,R_{\text{flex}}$ and transports it linearly. As the membrane bending rigiditiy is gradually decreased to $\kappa_H = 100\,k_BT$, the flexicle stops transporting the sphere and instead oscillates along its surface. 
  \item[Movie 8:] Two flexicles with $\kappa_H = 100\,k_BT$ sequentially latch onto a fixed spherical object with radius $R_{\text{sph}}=1.3\,R_{\text{flex}}$.  
  \item[Movie 9:] Multiple flexicles with $\kappa_H = 100\,k_BT$ sequentially latch to a fixed spherical object $R_{\text{sph}}=1.9\,R_{\text{flex}}$. 
  \item[Movie 10:] Multiple vesicles with bending rigidity $\kappa_H = 100\,k_BT$, subjected to constant force $F_G = 3\epsilon$ in the $-z$-direction, cooperate inside a square box by pushing one flexicle over a wall of height $h_{\text{wall}} = 1.5\,R_{\text{flex}}$. The wall height is too tall for a single vesicle to escape independently. 
  \item[Movie 11:] Multiple vesicles with bending rigidity $\kappa_H = 100\,k_BT$, subjected to constant force $F_G = 3\epsilon$ in the $-z$-direction, cooperate inside a square box by climbing on top of each other to overcome a wall of height $h_{\text{wall}} = 1.75\,R_{\text{flex}}$. The wall height is too tall for a single vesicle to escape independently 

\end{enumerate}
\pagebreak

\begin{figure}[ht]
  \centering
    \includegraphics[width = 0.8 \textwidth]{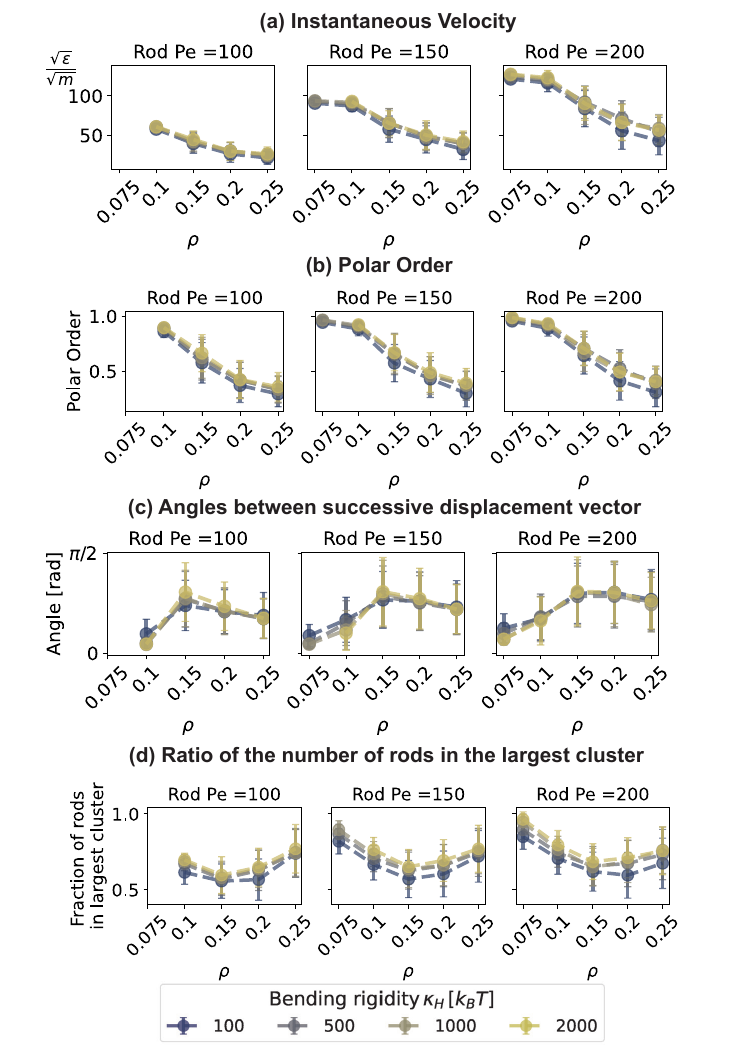}
    \caption{\label{fig:S1} Transport properties and collective alignment of self-propelled rods inside a single flexicle during propulsion at different rod densities $\rho$, mesh bending rigidities $\kappa_H$ and rod active velocities. Each category is presented in three columns corresponding to different P\'eclet numbers $\text{Pe} = 100$ (left), $\text{Pe} = 150$ (center), and $\text{Pe} = 200$ (right), plotted against $\rho$. (a) Mean instantaneous velocity of the flexicle centroid. (b) Polar order parameter of the rod particles. (c) Mean angle between successive instantaneous flexicle displacement directions separated by a timestep $\Delta t = 100\,\tau$, serving as a measure of angular correlation, analogous to rotational diffusion. (d) The ratio of the number of rod particles in the largest cluster.}
\end{figure}

\begin{figure}[ht]
  \centering
  \includegraphics[width = 0.95 \textwidth]{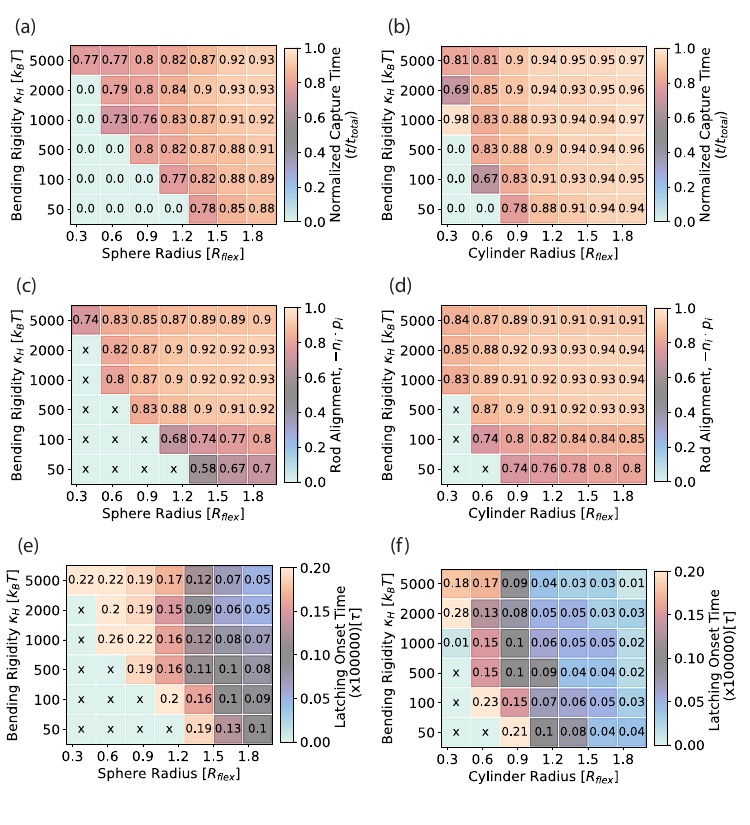}
  \caption{\label{fig:S2} Top panels show the ratio of the duration that a flexicle remains stably latched to the surface of a spherical (a) or cylindrical (b) object relative to the total time. Center panels illustrate the average alignment of the enclosed rod $n_i$ relative to the normal direction of the sphere (c) and cylinder (d) surface at the contact point $\hat{p}_i$. Bottom panels illustrate the latching onset time on the surface of a sphere (e) or cylinder (f). Each metric is plotted against the obstacle size and the membrane bending rigidity $\kappa_{H}$. All data shown correspond to simulations with a P\'eclet number $\text{Pe} = 100$.} 
  
\end{figure}

\begin{figure}[ht]
  \centering
  \includegraphics[width = 0.95 \textwidth]{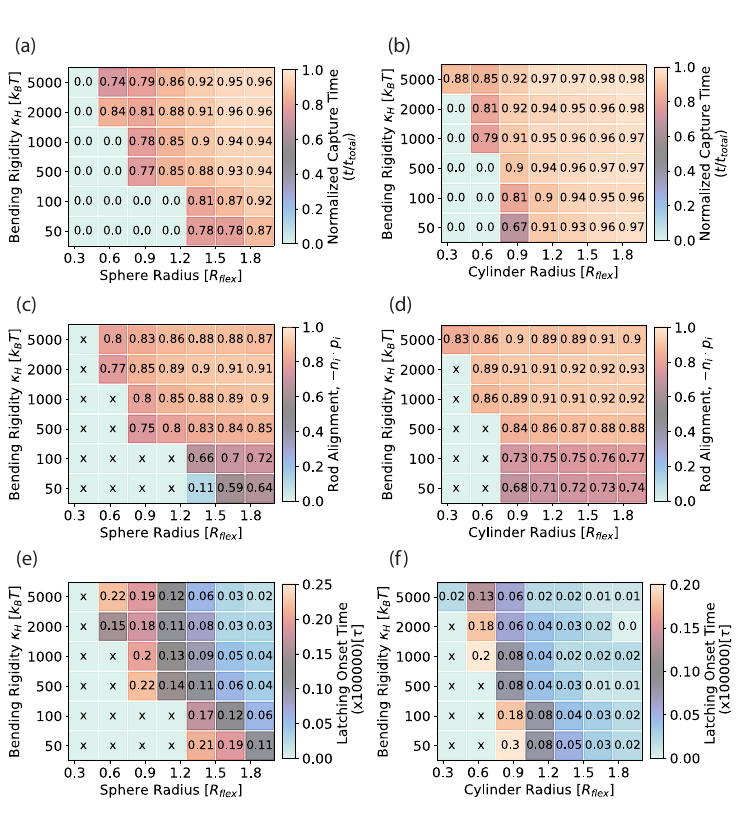}
  \caption{\label{fig:S3} Top panels show the ratio of the duration that a flexicle remains stably latched to the surface of a spherical (a) or cylindrical (b) object relative to the total time. Center panels illustrate the average alignment of the enclosed rod $n_i$ relative to the normal direction of the sphere (c) and cylinder (d) surface at the contact point $\hat{p}_i$. Bottom panels illustrate the latching onset time on the surface of a sphere (e) or cylinder (f). Each metric is plotted against the obstacle size and the membrane bending rigidity $\kappa_{H}$. All data shown correspond to simulations with a P\'eclet number $\text{Pe} = 200$.} 
  
\end{figure}

\begin{figure}[ht]
  \centering
  \includegraphics[width = 0.9 \textwidth]{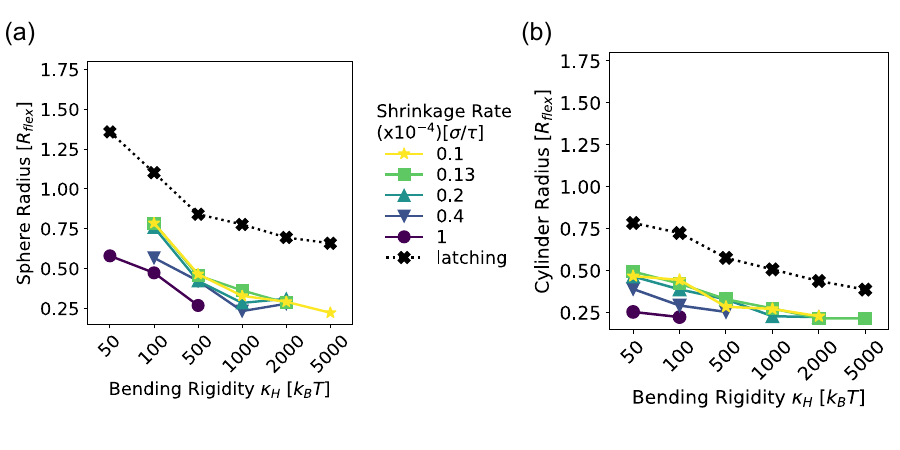}
  \caption{\label{fig:S4} Obstacle radius corresponding to a 50\% latching retention probability for a flexicle initially latched onto a shrinking (a) spherical or (b) cylindrical obstacle. The latching retention probability data were fitted with an error function of the form $f(x) = \frac{1}{2} \left(1 + \operatorname{erf}(a(x - b))\right)$ to determine the obstacle radius at which the flexicle has a 50\% probability of remaining latched. The resulting fitted radii are plotted as a function of membrane bending rigidity \( \kappa_H \). Each curve (color and marker) corresponds to a different shrinkage rate, defined by the rate of linear decrease in obstacle radius over time. The obstacle radius corresponding to 50\% latching success of Fig.\ref{fig:results1}(d) and Fig.\ref{fig:results3}(b) is also shown as reference (denoted by \xmark). All data shown correspond to simulations with a P\'eclet number $\text{Pe} = 100$.} 
  
\end{figure}

\begin{figure}[ht]
  \centering
  \includegraphics[width = 0.85\textwidth]{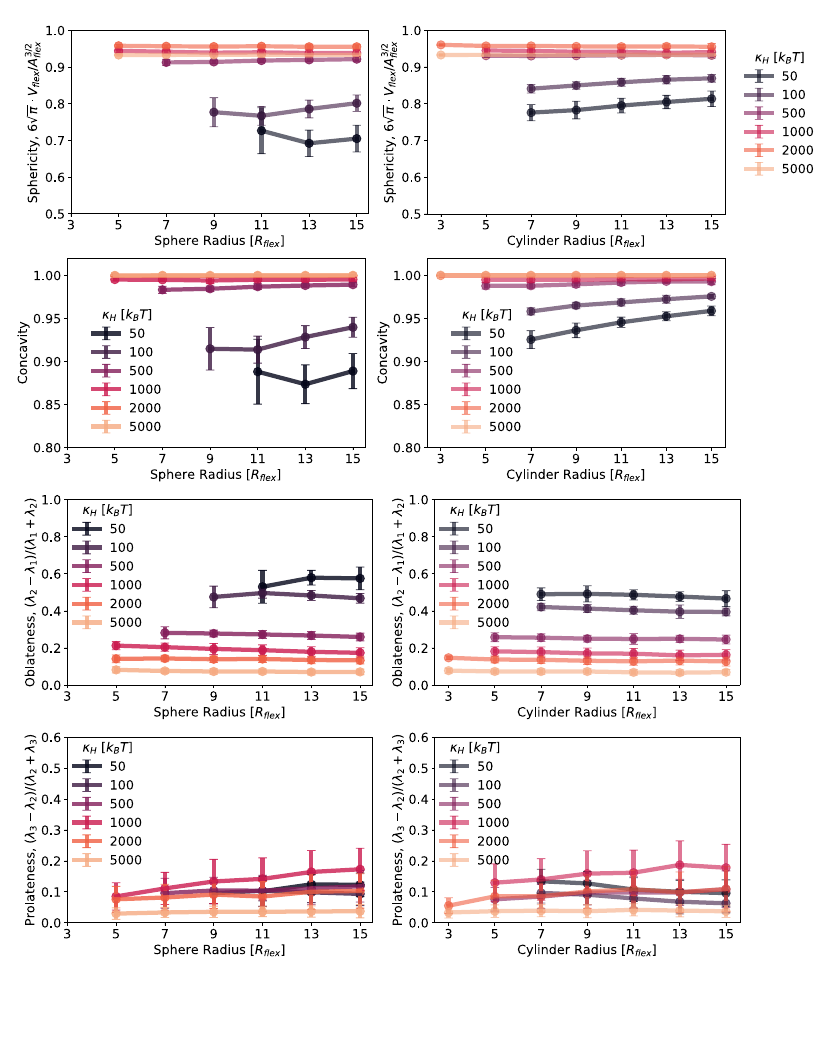}
  \caption{\label{fig:S5} Shape parameters of flexicles stably latched to the surfaces of spherical (left column) and cylindrical (right column) obstacles at different obstacle sizes and bending rigidities $\kappa_H$. The first row shows the sphericity, the second row shows concavity, measured by normalizing the flexicle volume by the volume of its convex hull and the third and fourth row display the oblateness and prolateness, defined as by the eigenvalues $\lambda_1 < \lambda_2 < \lambda_3$ of the flexicle’s gyration tensor. All data shown correspond to simulations with a P\'eclet number $\text{Pe} = 100$.}
  
\end{figure}

\begin{figure}[ht]
  \centering
  \includegraphics[width = 0.9 \textwidth]{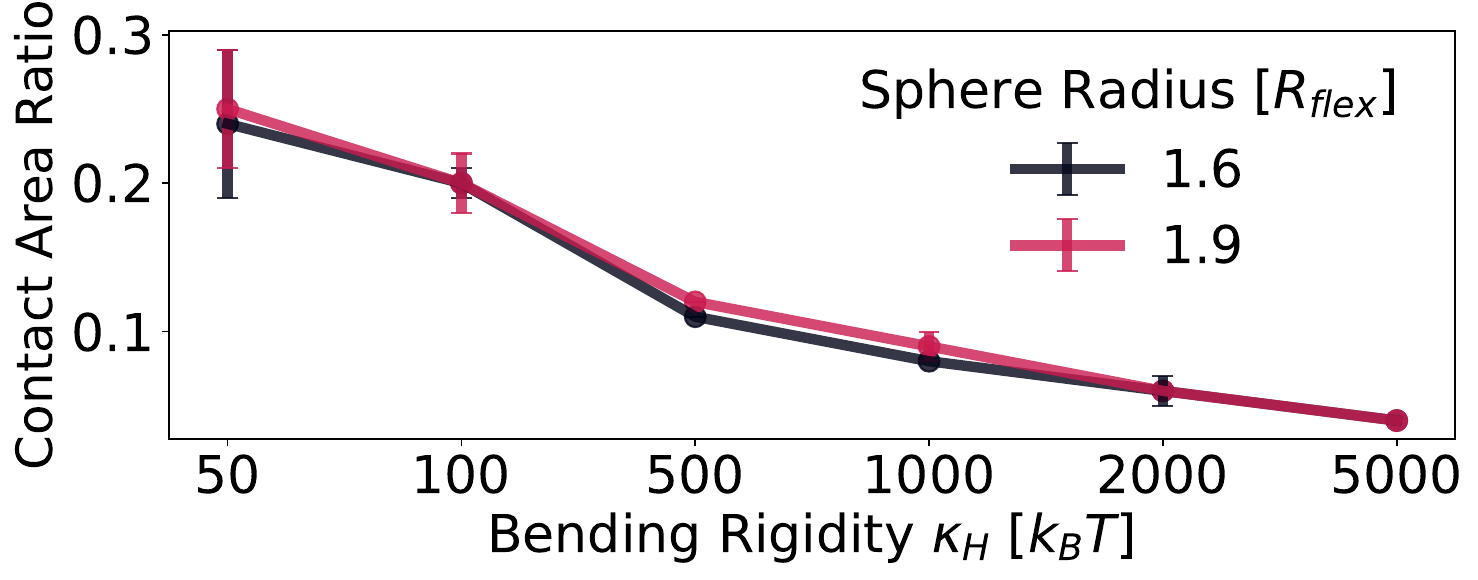}
  \caption{\label{fig:S6} The area of contact between the flexicle membrane and the sphere obstacle surface while latched as a function of mesh bending rigidity $\kappa_{H}$. All data shown correspond to simulations with a P\'eclet number $\text{Pe} = 100$.} 
  
\end{figure}

\begin{figure}[t!]
  \centering
  \includegraphics[width = 0.85 \textwidth]{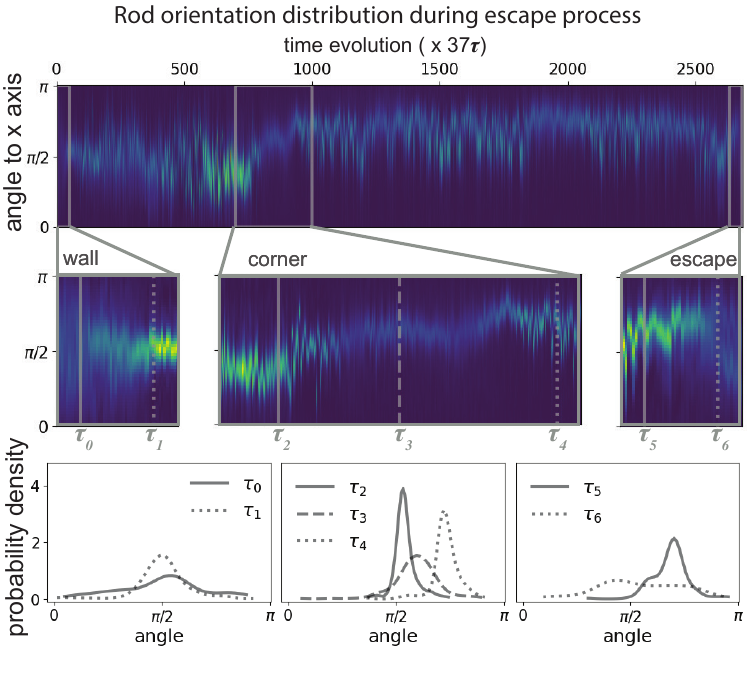}
  \caption{\label{fig:S7} Time evolution of the probability distribution of angles between the rods' long axes and the x-axis of the square wall under applied adhesion force grounding the flexicle to the floor. The first row shows the time evolution of these angle distributions. The second row contains three zoomed-in graphs (highlighted in gray boxes) from the first row, capturing critical moments: when the flexicle first contacts the wall, when it reaches the corner, and when it escapes the square wall. The third row provides a detailed comparison of the angle distributions at specific time points corresponding to each zoomed-in moment. The graphs in the third row depict the angle distributions at designated time points: $\tau_0$ and $\tau_1$, before and after the flexicle contacts the wall; $\tau_2$, $\tau_3$, and $\tau_4$, when the flexicle encounters the corner and the rods rearrange; and $\tau_5$ and $\tau_6$, before and after the flexicle escapes the square wall. Each of these time points is marked in the second row's zoomed-in graphs.
}
\end{figure}

\begin{figure}[t!]
  \centering
  \includegraphics[width = 0.85 \textwidth]{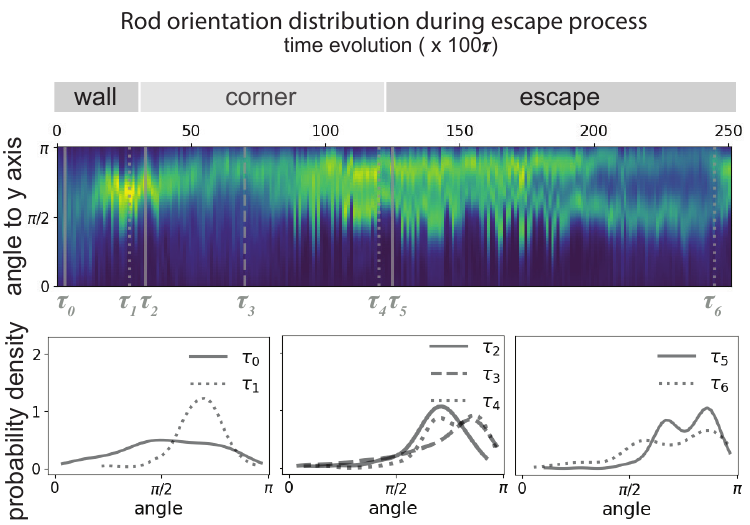}
  \caption{\label{fig:S8} Time evolution of the probability distribution of angles between the rods' long axes and the y-axis of the square wall , under a gravity-like force pressing the flexicle to the floor. The first row shows the time evolution of these angle distributions. The second row contains three zoomed-in graphs (highlighted in gray boxes) from the first row, capturing critical moments: when the flexicle first contacts the wall, when it reaches the corner, and when it escapes the square wall. The third row provides a detailed comparison of the angle distributions at specific time points corresponding to each zoomed-in moment. The graphs in the third row depict the angle distributions at designated time points: $\tau_0$ and $\tau_1$, before and after the flexicle contacts the wall; $\tau_2$, $\tau_3$, and $\tau_4$, when the flexicle encounters the corner and the rods rearrange; and $\tau_5$ and $\tau_6$, before and after the flexicle escapes the square wall. Each of these time points is marked in the second row's zoomed-in graphs.
}
\end{figure}

\end{document}